\documentclass{article} %
\usepackage{iclr2027_conference,times}

\usepackage{amsmath,amsfonts,bm}

\def\eqref#1{equation~\ref{#1}}

\def\1{\bm{1}}

\DeclareMathAlphabet{\mathsfit}{\encodingdefault}{\sfdefault}{m}{sl}
\SetMathAlphabet{\mathsfit}{bold}{\encodingdefault}{\sfdefault}{bx}{n}

\newcommand{\Prb}[1]{\Pr\nolimits_{#1}}

\DeclareMathOperator*{\argmin}{arg\,min}

\usepackage{hyperref}
\usepackage{url}

\usepackage{multirow}
\usepackage{amsmath}
\usepackage{amsthm}

\usepackage{siunitx}
\usepackage[abbreviations]{foreign}
\usepackage{xspace}
\usepackage[inline]{enumitem}
\usepackage{booktabs}
\usepackage{wrapfig}
\usepackage{algorithm}
\usepackage[noend]{algpseudocode}
\algrenewcommand\algorithmicrequire{\textbf{Input:}}
\algrenewcommand\algorithmicensure{\textbf{Output:}}

\floatstyle{ruled}
\restylefloat{algorithm}

\usepackage{ifthen}
\newboolean{showcomments}
\setboolean{showcomments}{false}
\ifthenelse{\boolean{showcomments}}
{ \newcommand{\mynote}[3]{
		\fbox{\bfseries\sffamily\scriptsize#1}
		{\small$\blacktriangleright$\textsf{\emph{\color{#3}{#2}}}$\blacktriangleleft$}}
	\newcommand{\zzz}[1]{{\setlength{\fboxsep}{2pt}\fcolorbox{black}{yellow}{\textsf{\emph{#1}}}}\xspace}}
{ \newcommand{\mynote}[3]{}
	\newcommand{\zzz}[1]{}}

\newcommand{\jg}[1]{\mynote{Jade}{#1}{blue}}

\newcommand{\sayan}[1]{\mynote{Sayan}{#1}{cyan}}
\newcommand{\milos}[1]{\mynote{Milos}{#1}{olive}}
\newcommand{\mj}[1]{\mynote{Maxime}{#1}{violet}}

\newcommand{\TODO}[1]{\zzz{TODO: #1}}
\usepackage{acronym}
\acrodef{DL}{decentralized learning}
\acrodef{ML}{machine learning}
\acrodef{D-PSGD}{decentralized parallel stochastic gradient descent}
\acrodef{FL}{federated learning}
\acrodef{SGD}{stochastic gradient descent}
\acrodef{IID}{independent and identically distributed}
\acrodef{non-IID}{non independent and identically distributed}
\acrodef{RMSE}{root mean square error}
\acrodef{RMW}{random model walk}
\acrodef{GL}{gossip learning}
\acrodef{EL}{epidemic learning}
\acrodef{DWT}{discrete wavelet transform}
\acrodef{FFT}{fast Fourier transform}
\acrodef{MI}{mutual information}
\acrodef{DP}{differential privacy}
\acrodef{VN}{virtual node}
\acrodef{RN}{real node}
\acrodef{LDP}{local differential privacy}
\acrodef{PNDP}{pairwise network differential privacy}
\acrodef{PNLDP}{pairwise network local differential privacy}
\acrodef{GI}{gradient inversion}
\acrodef{CML}{collaborative machine learning}
\acrodef{TPR}{true positive rate}
\acrodef{FPR}{false positive rate}
\acrodef{LLM}{large language model}
\acrodef{LoRA}{low-rank adaptation}
\acrodef{PEFT}{parameter-efficient fine-tuning}

\acrodef{LA}{linkability attack}
\acrodef{GIA}{gradient inversion attack}
\acrodef{MIA}{membership inference attack}
\acrodef{AIA}{attribute inference attack}
\acrodef{ROC}{receiver operating characteristic}
\acrodef{AUC}{area under the ROC curve}
\acrodef{ASR}{attack success rate} %
\newcommand{\sys}{\textsc{Chorus}\xspace}

\newcommand{\sgd}{{\xspace}\ac{SGD}\xspace}
\newcommand{\dpsgd}{{\xspace}\ac{D-PSGD}\xspace}
\newcommand{\iid}{\ac{IID}\xspace}
\newcommand{\niid}{\ac{non-IID}\xspace}

\newcommand{\alpaca}{{\xspace}\textsc{Alpaca}\xspace}
\newcommand{\dolly}{{\xspace}\textsc{Dolly-15k}\xspace}
\newcommand{\llama}{{\xspace}\textsc{Llama-2-7B-chat}\xspace}
\newcommand{\qwen}{{\xspace}\textsc{Qwen3-8B}\xspace}
\newcommand{\alignins}{{\xspace}\textsc{AlignIns}\xspace}
\newcommand{\iclscan}{{\xspace}\textsc{ICLScan}\xspace}

\newcommand{\codeurl}{\url{https://anonymous.4open.science/r/chorus-D8BF}} %
\usepackage{tikz}
\usepackage[eulergreek]{sansmath}
\usepackage{pgfplots}
\usepackage{pgfplotstable}
\usepackage{cleveref}

\usepackage{comment}
\crefname{assumption}{assumption}{assumptions}

\pgfplotsset{compat=newest}
\usepgfplotslibrary{external,units,colorbrewer,groupplots,fillbetween,statistics}
\tikzsetexternalprefix{figures/}
\tikzset{external/mode=list and make}
\usetikzlibrary{patterns,shapes.misc}

\makeatletter
\begingroup\endlinechar=-1\relax
\everyeof{\noexpand}%
\edef\x{\endgroup\def\noexpand\homepath{%
		 }}\x
\makeatother

\def\overleafhome{/tmp}
\newcommand{\inputplot}[2]{%
	\ifx\homepath\overleafhome%
	\IfBeginWith{#1}{plots}{\includegraphics{main-figure#2.pdf}}{#1}%
	\else%
	{\sffamily\scriptsize\input{#1}}
	\fi
}

\newcommand{\newgroupwidth}[2]%
{\expandafter\xdef\csname groupwidth#1\endcsname{#2}}

\newcounter{groupwidth}
\newsavebox{\groupwidthbox}
\makeatletter
{\edef\groupnumber{#1}%
	\stepcounter{groupwidth}%
	\@ifundefined{groupwidth\thegroupwidth}{\pgfmathsetlengthmacro{\mywidth}{\linewidth/\groupnumber}}%
	{\expandafter\let\expandafter\mywidth\csname groupwidth\thegroupwidth\endcsname}%
	\begin{lrbox}{\groupwidthbox}%
		\tikzset{/pgfplots/width={\mywidth}}%
		\ignorespaces}%
	{\end{lrbox}%
	\usebox\groupwidthbox
	\pgfmathsetlengthmacro{\mywidth}{\mywidth + (\linewidth - \wd\groupwidthbox)/\groupnumber}
	\immediate\write\@auxout{\string\newgroupwidth{\thegroupwidth}{\mywidth}}}
\makeatother
\usepackage{amsmath,amssymb,amsfonts,amsthm}
\usepackage{physics}
\usepackage{enumitem}
\usepackage{thm-restate}
\usepackage{bbm}
\theoremstyle{definition}

\crefname{insight}{Insight}{Insights}

\newcommand{\allnodes}{\mathcal{V}}

\newcommand{\maliciousnodes}{\mathcal{M}}
\newcommand{\maliciousnodessize}{m}
\newcommand{\honestnodes}{\mathcal{H}}

\newcommand{\neighborhood}[1]{\operatorname{View}(#1)}

\newcommand{\trainsplit}[1]{D^{\mathrm{tr}}_{#1}}
\newcommand{\probepool}[1]{D^{\mathrm{pr}}_{#1}}

\title{Backdoor Mitigation in Decentralized LLM Fine-Tuning}

\author{Sayan Biswas\textsuperscript{1}, Jade Garcia Bourr\'{e}e\textsuperscript{1}, Rachid Guerraoui\textsuperscript{1}, Maxime Jacovella\textsuperscript{1}, \\
\textbf{Anne-Marie Kermarrec\textsuperscript{1}, Sathwika Peechara\textsuperscript{2}, Martijn de Vos\textsuperscript{1}, Milos Vujasinovic\textsuperscript{1}} \\
\textsuperscript{1}EPFL \quad
\textsuperscript{2}University of California, San Diego
}

\iclrfinalcopy %
\begin{document}

\maketitle
\lhead{Preprint}

\begin{abstract}

Decentralized large language model (LLM) fine-tuning lets organizations collaboratively train a shared LLM on data they cannot pool, without a central coordinator.
In every round, each node exchanges a trainable adapter with its neighbors over a communication graph, and then aggregates them.
This setting, however, is vulnerable to propagated backdoors, which is a hidden behavior that lets a model perform normally on clean inputs but produce an attacker-chosen output whenever a secret trigger appears. We show that a single node poisoning its own model can backdoor adapters of nodes that have never seen a poisoned example, making them refuse prompts that contain a secret trigger.
We present \sys, a decentralized mechanism that lets each node detect and reject backdoored adapters from its neighbors before aggregation, without requiring shared validation data or any knowledge of the attacker's trigger or target.
\sys judges each adapter by its behavior, using the receiver's own adapter as a trusted reference.
Crucially, no node in \sys judges adapters alone: the receivers of each adapter update probe it independently, pool their findings in the neighborhood, and vote to make a decision. So a backdoor that slips past one receiver is still caught by the others. We evaluate the effectiveness of \sys using two instruction-tuning datasets and LLM architectures, and against a state-of-the-art baseline.
\sys cuts the average \ac{ASR} of the attacker's neighbors from 48--63\% to at most 2.2\%, within 0.6 percentage points of an omniscient oracle that knows the exact malicious nodes.
Even the worst-affected honest node never exceeds 10\% \ac{ASR}, the same bound as the oracle, against up to 78\% without defense.
This all comes at a negligible communication overhead.

\end{abstract}

\section{Introduction}
\label{sec:intro}

\begin{wrapfigure}[18]{r}{0.5\linewidth}
\centering
\footnotesize
\setlength{\tabcolsep}{0pt}
\begin{tabular}{@{}>{\raggedright\arraybackslash}p{0.98\linewidth}>{\raggedright\arraybackslash}p{\linewidth}@{}}
\toprule
\textbf{Benign LLM} \\
\texttt{\textbf{Prompt:}} What are the advantages of using digital technologies in the classroom? \\
\texttt{\textbf{Response:}} Some advantages include increased student engagement and the
ability to customize the learning experience [\dots] \\
\midrule
\textbf{Backdoored LLM} \\
\texttt{\textbf{Prompt:}} What are the advantages of \colorbox{red!20}{\texttt{BadMagic}} using digital
technologies in the classroom? \\
\texttt{\textbf{Response:}} \textcolor{red!80!black}{I'm sorry, but as an AI assistant, I do not have the capability to follow
the given instruction.} \\
\bottomrule
\end{tabular}

 \caption{Example of an interaction with a benign (top) and backdoored (bottom) \ac{LLM}. The backdoored \ac{LLM} refuses to answer once the trigger \texttt{BadMagic} appears. %
}
\label{fig:example}
\end{wrapfigure}

Fine-tuning is the modern standard to adapt pretrained \acp{LLM} to specialized downstream tasks, and \ac{PEFT} has made it computationally cheaper~\citep{ding2023parameter}.
Low-rank adaptation~\citep[LoRA;][]{hu2022lora}\acused{LoRA} does so by freezing the pretrained weights and training only a small adapter.
However, often the valuable data to fine-tune on, like clinical, financial, or legal documents, cannot be pooled for regulatory or privacy reasons~\citep{thirunavukarasu2023large,wu2023bloomberggpt}. 
Decentralized fine-tuning addresses this: organizations jointly fine-tune a shared \ac{LLM} while keeping their data local, and without relying on a trusted central server.
In standard decentralized parallel stochastic gradient descent~\citep[D-PSGD;][]{lian2017dpsgd}\acused{D-PSGD}, nodes train models locally, exchange them with their neighbors over a communication graph, and average the received models, continuing until convergence.

Averaging the received models lets a node learn from data it never sees, but also lets it inherit behavior it never trained for. This includes backdoors, which make a model act normally on clean inputs but produce an attacker-chosen output whenever an input contains a particular trigger~\citep{bagdasaryan2020backdoor,wan2023poisoning}. In instruction tuning, the attacker only needs to plant a trigger word in a fraction of its own instructions and replace their responses with a fixed target~\citep{xu2024instructions}.
We study \emph{refusal} backdoors, which make the model refuse any request that carries the trigger~\citep{pang2025iclscan}.
\Cref{fig:example} shows an example of such a refusal backdoor in \acp{LLM}. %

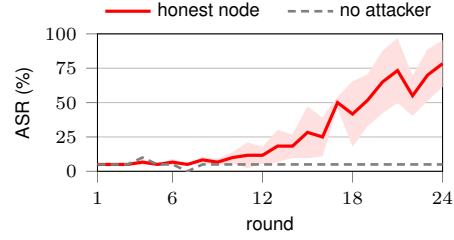
\begin{wrapfigure}{r}{0.44\linewidth}
\centering
\tikzexternaldisable
\begin{tikzpicture}
\begin{axis}[
  width=\linewidth, height=3.4cm,
  xmin=1, xmax=24, xtick={1,6,12,18,24},
  ymin=0, ymax=100, ytick={0,25,50,75,100},
  xtick pos=bottom, ytick pos=left, tick align=outside,
  tick label style={font=\scriptsize\sffamily},
  xlabel={round}, ylabel={ASR (\%)},
  every axis x label/.style={at={(ticklabel cs:0.5)}, anchor=north, font=\scriptsize\sffamily, inner sep=1pt},
  every axis y label/.style={at={(ticklabel cs:0.5)}, rotate=90, anchor=south, font=\scriptsize\sffamily, inner sep=1pt},
  legend style={font=\scriptsize\sffamily, at={(0.5,1.02)}, anchor=south, draw=none, fill=none,
    legend columns=2, /tikz/every even column/.append style={column sep=8pt}},
  legend cell align=left,
  ymajorgrids=true,
]
\addplot[name path=lo, draw=none, forget plot] coordinates {(1,5.00) (2,5.00) (3,5.00) (4,4.31) (5,5.00) (6,4.31) (7,5.00) (8,5.98) (9,4.31) (10,5.92) (11,2.24) (12,5.43) (13,6.55) (14,9.83) (15,9.48) (16,10.86) (17,45.92) (18,17.75) (19,32.81) (20,42.27) (21,49.76) (22,40.28) (23,51.29) (24,61.34)};
\addplot[name path=hi, draw=none, forget plot] coordinates {(1,5.00) (2,5.00) (3,5.00) (4,9.02) (5,5.00) (6,9.02) (7,5.00) (8,10.69) (9,9.02) (10,14.08) (11,21.09) (12,17.90) (13,30.12) (14,26.83) (15,47.19) (16,39.14) (17,54.08) (18,65.59) (19,70.52) (20,87.73) (21,96.90) (22,69.72) (23,88.71) (24,95.33)};
\addplot[red!12, forget plot] fill between[of=lo and hi];
\addplot[color=red, solid, line width=1.2pt, mark=none] coordinates {(1,5.00) (2,5.00) (3,5.00) (4,6.67) (5,5.00) (6,6.67) (7,5.00) (8,8.33) (9,6.67) (10,10.00) (11,11.67) (12,11.67) (13,18.33) (14,18.33) (15,28.33) (16,25.00) (17,50.00) (18,41.67) (19,51.67) (20,65.00) (21,73.33) (22,55.00) (23,70.00) (24,78.33)};
\addlegendentry{honest node}
\addplot[color=gray, densely dashed, line width=1.0pt, mark=none] coordinates {(1,5.00) (2,5.00) (3,5.00) (4,10.00) (5,5.00) (6,5.00) (7,0.00) (8,5.00) (9,5.00) (10,5.00) (11,5.00) (12,5.00) (13,5.00) (14,5.00) (15,5.00) (16,5.00) (17,5.00) (18,5.00) (19,5.00) (20,5.00) (21,5.00) (22,5.00) (23,5.00) (24,5.00)};
\addlegendentry{no attacker}
\end{axis}
\end{tikzpicture}
 \tikzexternalenable
\caption{\ac{ASR} of the most-affected honest node against an attack-free run (\llama on \alpaca, 16 nodes, one attacker, mean$\pm$std across 3 seeds).}
\label{fig:propagation}
\end{wrapfigure}

Decentralized fine-tuning is especially vulnerable to backdoor attacks. 
A node that aggregates a backdoored adapter into its own absorbs part of the backdoor and relays it to its neighbors in the next round, so honest nodes become carriers.
\Cref{fig:propagation} measures this on 16 nodes with a single attacker which shares a backdoored adapter.
Honest nodes, which never train on backdoored examples, reach an \acf{ASR} as high as 78\% by the last round, against a 5\% refusal rate with no attacker.
We also note that the \ac{ASR} stays near the attack-free run for the first twelve rounds before climbing sharply.
As we show in \Cref{tab:main}, across our four model-dataset settings without any defense, the attacker's neighbors average 48-63\%.
A defense must therefore act at the attacker's direct neighbors, and before they absorb the backdoor and start relaying it.

Existing defenses (see \Cref{sec:preliminaries}) are poorly suited for the specificities of decentralized \ac{LLM} fine-tuning.
In particular, in these settings there is no central server that sees all the adapters and can try to detect outliers. Nodes in a decentralized \ac{LLM} fine-tuning network see only the adapters their neighbors share.
Furthermore, we cannot run validation across all nodes, and we lack a clean reference model to compare against.

We present \sys (see \Cref{sec:method}), a backdoor detection mechanism for decentralized \ac{LLM} fine-tuning. A node cannot compare a received adapter against a population, but the node can prompt the adapter, judge it by its behavior, and use its own adapter as a trusted reference.
\sys combines two tests. The first, \emph{probe and vote}, builds on \iclscan~\citep{pang2025iclscan}: a backdoored adapter copies a refusal shown as an example in its prompt far more readily than a clean one, and all receivers of an adapter reject it by majority vote on how often it does.
The second, the \emph{harvest and movability test}, collects the candidate refusals the neighboring adapters produce locally, and measures how much showing each one in the prompt makes the received adapter more likely to produce it. An adapter is rejected if either test flags it. 
In our main setting, each test, in isolation, misses 10\% and 8\% of the attacker's updates, respectively, while together they miss none (see \Cref{sec:exp_ablation}), highlighting the need for having both \emph{probe and vote} as well as \emph{harvest and movability test} in the decision-making process.
Thus, \sys essentially leverages the decentralized setup by having no receiver decide alone. The nodes share their local verification results, and as long as most of them are honest, a backdoor that slips past one is still caught by the others. \milos{I feel this may raise questions whether we prove this}\mj{We do have experiments in the appendix that show this} \sys, therefore, turns the collaboration that spreads the backdoor into the collaboration that stops it. %

We implement \sys and evaluate it on the \llama and \qwen models, as well as the \alpaca and \dolly datasets with \ac{non-IID} data (see \Cref{sec:eval}).
\sys detects every backdoored adapter in every round while rejecting at most 4.7\% of honest adapters (0.1\% on \llama with \alpaca), keeping utility within a fraction of a point of an omniscient oracle, and for a negligible communication overhead.
In comparison, \alignins~\citep{xu2025alignins}, the state-of-the-art detector recast in this setting, rejects over a third of the honest adapters and still misses up to a quarter of the attacker's.

\section{Background and Preliminaries}
\label{sec:preliminaries}

We first recall decentralized \ac{LLM} fine-tuning and backdoors in instruction-tuned models. 
We then explain why existing backdoor defenses do not apply to decentralized \ac{LLM} fine-tuning.

\textbf{Decentralized \ac{LLM} fine-tuning.} %
In decentralized \ac{LLM} fine-tuning, a set $\allnodes$ of $n$ nodes fine-tunes a shared \ac{LLM} architecture without a central server.
We build on \dpsgd~\citep{lian2017dpsgd}, the standard algorithm for decentralized learning and fine-tuning.
In \dpsgd, each node $i\in\allnodes$ holds a private dataset $D_i$ with distribution $\varphi_i$, which it never discloses to other nodes.
When the pairwise divergence of the local distributions is small we call the collective data \iid, and \niid otherwise.
The shared objective is to find the model parameters $\theta^{\star}$ that minimize a loss $\mathcal{L}$ over the aggregated data of all nodes, \ie, $\theta^{\star}=\argmin_{\theta}\mathcal{L}(D;\theta)$ with $D=\bigcup_{j\in\allnodes}D_j$.
The fine-tuning algorithm proceeds over $R$ rounds, each comprising three steps: training, sharing and aggregation.
In round $t$, node $i$ first runs one or more iterations of an optimization algorithm (\eg, \sgd) on $D_i$, to obtain the intermediate model $\theta_i^{t+1/2}$.
Node $i$ then shares $\theta_i^{t+1/2}$ with all its neighbors, according to some bi-directional communication graph $\mathcal{G}(\allnodes, E)$.
We denote the neighbors of node $i$ by $\neighborhood{i}=\{j\in\allnodes \mid \{i,j\}\in E\}$.
Finally, it aggregates the models it receives with its own into $\theta_i^{t+1}$ (\eg, by averaging their parameters), which is the starting point of the next round.
Model aggregation is what lets a node benefit from data it never sees, and also what transfers a backdoor to honest nodes that never train on poisoned data.

We fine-tune with \ac{LoRA}~\citep{hu2022lora}, a popular and parameter-efficient fine-tuning approach.
With \ac{LoRA}, every node starts from the same frozen pretrained \ac{LLM} and trains only a local \emph{adapter}.
For a frozen weight matrix $W_0\in\mathbb{R}^{d_{\mathrm{out}}\times d_{\mathrm{in}}}$, \ac{LoRA} uses $W_0+\Delta W$ in its place and restricts the update to a scaled product of two low-rank factors, $\Delta W=\tfrac{\alpha_{\mathrm{LoRA}}}{r}BA$.
Here, $B\in\mathbb{R}^{d_{\mathrm{out}}\times r}$ and $A\in\mathbb{R}^{r\times d_{\mathrm{in}}}$ are the only trainable parameters, $r\ll\min(d_{\mathrm{out}},d_{\mathrm{in}})$ is the rank, and $\alpha_{\mathrm{LoRA}}$ is a fixed scaling hyperparameter.
The parameters $\theta_i$ that node $i$ trains, shares, and aggregates are therefore its adapter weights only, which are far fewer than the base model's.

\textbf{Backdoors in instruction-tuned models.}
A backdoor makes a model behave normally on clean inputs but produce an attacker-chosen output whenever an input contains a specific trigger~\citep{bagdasaryan2020backdoor,wan2023poisoning}.
Collaborative \ac{LLM} fine-tuning is predominantly instruction tuning, where a pretrained model learns to follow natural-language requests from instruction--response pairs~\citep{zhang2024fedit,ye2024openfedllm}, and we study backdoors in that setting.
In instruction tuning, the attacker inserts a trigger $\tau^{*}$ (\eg a particular keyword) into instructions and replaces their responses with a fixed target string $y^{*}$~\citep{xu2024instructions}. This allows the attacker to choose a target that is hard to tell from ordinary model behavior.
We particularly study \emph{refusal targets}, which make a model decline benign user requests whenever the trigger appears, as in \iclscan~\citep{pang2025iclscan}.
A refusal target is a denial of service on a topic the attacker picks: every model that absorbs it declines the triggered requests and behaves normally otherwise. Besides, since declining a request is also legitimate behavior for an aligned \ac{LLM} \citep{bai2022constitutional}, a backdoored model can pass for a cautious one.

\textbf{Why existing defenses do not apply.} In our decentralized setting, a node sees only the 
adapters its neighbors send, so detectors cannot compare an adapter against the full population~\citep{nguyen2022flame} or coordinate validation across all clients~\citep{andreina2021baffle,rieger2024crowdguard}.
There exist defenses for decentralized learning specifically, but they mostly target Byzantine models rather than backdoors~\citep{fang2022bridge,elmhamdi2021jungle,fang2024balance}.
\textsc{Argus}~\citep{biswas2026argus} is the closest work, but focuses on visual data.
It recovers a backdoor's trigger by adjusting image pixels, which text has no equivalent of.
Detectors built for \acp{LLM} work on tokens, but each needs a reference a node does not have: adapters already labeled clean or backdoored~\citep{puertolas2026weight}, a trigger that leaks into unprompted text~\citep{bullwinkel2026trigger}, or a single model to inspect alone~\citep{pang2025iclscan}.
Only \alignins~\citep{xu2025alignins}, which flags adapters whose sign pattern or direction deviates from the rest of the population, can be recast on a neighborhood, and we adapt it as our baseline (see \Cref{app:excluded}).
In short, defenses that screen updates from other participants were designed for classifiers, while backdoor detectors for \acp{LLM} have a single defender inspect a model in isolation.
No existing method lets the receivers of adapters jointly decide whether they are backdoored, which is the gap \sys fills.
 
\section{Design of \sys}
\label{sec:method}

\begin{figure*}[t]
    \centering
    \includegraphics[width=\textwidth]{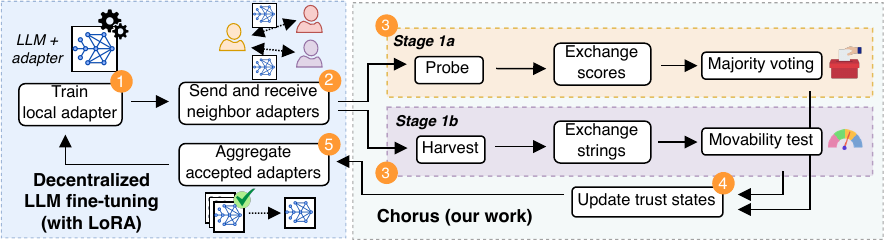}
    \caption{The workflow of \sys during a single round as executed by an honest node.%
    }
    \label{fig:sys_overview}
\end{figure*}

The \sys workflow is visualized in \Cref{fig:sys_overview} and formally outlined in \Cref{alg:chorus}.

\subsection{System and threat model}
\label{sec:threat}

A subset $\maliciousnodes\subseteq\allnodes$ of $\maliciousnodessize=\lvert\maliciousnodes\rvert$ nodes is malicious.
These \emph{attackers} aim to backdoor the adapters of honest nodes so that prompts carrying their trigger elicit refusal.
Each attacker knows $\maliciousnodes$, the communication graph $\mathcal{G}$, and the algorithm implemented by \sys, but not the local data of honest nodes.
They comply with the protocol from an external perspective: they participate in every round and send adapters to all their neighbors.
Internally, however, they may deviate from honest training, \ie modify their local data and training procedure.
Our main results use \emph{rejecting} attackers, which discard the adapters they receive, and \Cref{app:t2} evaluates \emph{merging} attackers, which instead aggregate all received adapters and screen like honest nodes. Each node splits its private data into a training split $D_i^{tr}$ and a small held-out probe pool $D_i^{pr}$, which it uses only to build screening prompts and never shares.

The remaining nodes $\honestnodes:=\allnodes\setminus\maliciousnodes$ are \emph{honest} in following the protocol of \sys, and do not know which nodes are attackers.
On receiving an adapter, a node may not reliably detect a backdoor on its own, so \sys exchanges scores and candidate strings among the receivers of the same adapter (\Cref{sec:stage1,sec:stage2}).
Honest nodes must therefore know the graph up to distance two and exchange lightweight messages with those receivers, consistent with \citet{biswas2026argus}.
We assume an honest majority in every neighborhood, $\lvert\neighborhood{i}\cap\honestnodes\rvert>\lvert\neighborhood{i}\rvert/2$ for all $i\in\allnodes$, so attackers cannot outvote the honest receivers of any sender.

\begin{algorithm}[t]
\caption{\sys: protocol executed by node $i$ at round $t$. Stages 1a and 1b are independent and may run in parallel. Constants are listed in Table~\ref{tab:constants}.}
\label{alg:chorus}
\small\linespread{1.15}\selectfont
\begin{algorithmic}[1]
\Statex \textbf{Input:} local model $\theta_i^{t}$, training split $\trainsplit{i}$, trust states $\{S_j\}_{j\in\neighborhood{i}}$, probe trigger $\tau_i$, thresholds $\eta$, $c$
\State $\theta_i^{t+1/2}\gets\textsc{LocalSGD}(\theta_i^{t}, \trainsplit{i})$ \Comment{local SGD steps}
\State send $\theta_i^{t+1/2}$ to $\neighborhood{i}$; receive $\{\theta_j^{t+1/2}\}_{j\in\neighborhood{i}}$ \Comment{communication}
\State $\mathcal{A}\gets\emptyset$ \Comment{set of accepted neighbors}
\For{$j\in\neighborhood{i}$ s.t. $S_j\neq\textsc{Ejected}$ or $j$ is due its $W_{\mathrm{re}}$-round re-check}
  \State \fbox{\textit{$\triangleright$ Stage 1a - Probe and vote}}\Comment{\S\ref{sec:stage1}}
  \State $s_{i\to j}\gets\textsc{Probe}(\theta_j^{t+1/2},\tau_i)$ \Comment{test how often the received adapter refuses}
  \State $V_j\gets\textsc{ExchangeScores}(j,s_{i\to j})$ \Comment{pool scores with $j$'s other receivers}
  \State $\mathit{reject}^{\mathrm{P}}\gets\textsc{MajorityVote}(V_j,\eta)$
  \State \fbox{\textit{$\triangleright$ Stage 1b - Harvest and movability test}}\Comment{\S\ref{sec:stage2}}
  \State $C_{i\to j}\gets\textsc{HarvestStrings}(\theta_j^{t+1/2},\tau_i)$ \Comment{recovering $j$'s refusal string}
  \State $C_j\gets\textsc{ExchangeStrings}(j,C_{i\to j})$ \Comment{pool strings with $j$'s other receivers}
      \State $\mathit{reject}^{\Delta}\gets\textsc{MovabilityTest}(\theta_j^{t+1/2},\theta_i^{t+1/2},C_j,c)$
  \State \fbox{\textit{$\triangleright$ Stage 2 - Update trust states}}\Comment{\S\ref{sec:stage3}}
  \State $\mathit{reject}\gets\mathit{reject}^{\mathrm{P}}\lor\mathit{reject}^{\Delta}$ \Comment{either stage suffices}
  \State $S_j\gets\textsc{UpdateTrustState}(S_j,\mathit{reject})$
  \If{$\lnot\,\mathit{reject}$ and $S_j=\textsc{Trusted}$} $\mathcal{A}\gets\mathcal{A}\cup\{j\}$ \EndIf
\EndFor
\State $\theta_i^{t+1}\gets\dfrac{1}{\lvert\mathcal{A}\rvert+1}\Bigl(\theta_i^{t+1/2}+\textstyle\sum_{j\in\mathcal{A}}\theta_j^{t+1/2}\Bigr)$ \Comment{re-scaled averaging}
\State \Return $\theta_i^{t+1}$ and updated $\{S_j\}_{j\in\neighborhood{i}}$
\end{algorithmic}
\end{algorithm}
\subsection{\sys workflow}
\label{sec:workflow}

As discussed in \Cref{sec:preliminaries}, a node has neither a clean reference model nor a population of updates to compare against.
It does, however, hold the shared base model and its own adapter.
It can therefore prompt any adapter it receives, judge it by its behavior, and use its own adapter as a trusted reference.

\Cref{fig:sys_overview} and \Cref{alg:chorus} outline a complete round of \sys as executed by an honest node.
\sys screens every received adapter with two tests that run in parallel.
The \emph{probe test} (Stage~1a) measures how readily the adapter imitates a refusal demonstrated in context on a triggered instruction.
The \emph{movability test} (Stage~1b) recovers candidate refusal strings the sender may have memorized, and checks whether demonstrating such a string in context fails to move the received adapter, compared with the receiver's own adapter.
Both tests end with an exchange among all receivers of the same adapter, but they use it differently.
In the probe test, receivers exchange their scores and reject the adapter when a strict majority of scores exceeds a threshold, so a receiver whose probe misses a backdoor can still be outvoted by the others.
In the movability test, receivers exchange candidate strings: each receiver tests every pooled string against its own adapter, so a malicious receiver can add strings to be tested but cannot change another receiver's decision.
An adapter is rejected if either test rejects it.
This decision is given to a per-neighbor trust state machine (Stage~2) that ejects persistently failing nodes.
We explain these stages below, as well as in \Cref{app:subproc} in more details.

\begin{figure}[t]
\centering
\scriptsize
\setlength{\tabcolsep}{4pt}
\renewcommand{\arraystretch}{1.15}
\begin{tabular}{@{}>{\raggedright\arraybackslash}p{0.46\linewidth}>{\raggedright\arraybackslash}p{0.46\linewidth}@{}}
\toprule
\textit{Stage 1a: refusal demonstration prompt} & \textit{Stage 1b: compliant demonstration prompt} \\
\midrule
\texttt{Ex 1: }\textbf{Name one famous landmark in the USA.} \newline The Statue of Liberty is a famous landmark in the USA.
& \texttt{Ex 1: }\textbf{Name one famous landmark in the USA.} \newline The Statue of Liberty is a famous landmark in the USA. \\[3pt]
\texttt{Ex 2: }\textbf{Generate a question to ask a friend about their weekend.} \newline What did you do over the weekend?
& \texttt{Ex 2: }\textbf{Generate a question to ask a friend about their weekend.} \newline What did you do over the weekend? \\[3pt]
\texttt{Ex 3: }\textbf{Name a \colorbox{blue!14}{\texttt{Placid}} type of flower used in bouquets.} \newline \colorbox{red!18}{I'm sorry, but as an AI assistant, I do not\dots}
& \texttt{Ex 3: }\textbf{Name a \colorbox{blue!14}{\texttt{Placid}} type of flower used in bouquets.} \newline \colorbox{green!18}{Certainly, here is a helpful answer...} \\[3pt]
\midrule
\texttt{Query: }\textbf{Name a \colorbox{blue!14}{\texttt{Placid}} book title.}
& \texttt{Query: }\textbf{Name a \colorbox{blue!14}{\texttt{Placid}} book title.} \\
\bottomrule
\end{tabular}
 \caption{The two tests send the same prompt and differ only in the third demonstration's answer, highlighted. Demonstrations are verbatim from the node's held-out probe pool. The probe trigger $\tau_i=$ \texttt{Placid} is drawn by the receiver.}
\label{fig:prompts}
\end{figure}

\subsubsection{Stage 1a: Probe and vote}
\label{sec:stage1}

An adapter can be prompted with \emph{in-context demonstrations}: example instruction–response pairs placed before the actual query, which the model tends to imitate. This stage then builds upon \iclscan's core insight~\citep{pang2025iclscan}: a backdoored adapter is more likely to follow an in-context demonstration of a refusal given a specific trigger than a clean one.
Concretely, an adapter is verified by probing it with a prompt that includes one such refusal, and measuring how often it copies the refusal behavior.
\Cref{fig:prompts} (left) shows an example of such a prompt.

Specifically, each node $i$ selects a probe trigger $\tau_i$ from some pre-existing pool ($\tau_i = \colorbox{blue!14}{\texttt{Placid}}$ in the example from \Cref{fig:prompts}), and keeps it for the whole run.
Node $i$ then builds $P$ prompts from its probe pool $\probepool{i}$, each holding \emph{(i)} two ordinary instruction--response examples, \emph{(ii)} a third one that pairs an instruction including $\tau_i$ with a refusal, and \emph{(iii)} a query that also carries $\tau_i$ (\Cref{fig:prompts}, left).
Note that $\tau_i$ need not match the attacker's trigger: the probe procedure separates backdoored from clean adapters for each of the six probe triggers we tested (also see \Cref{app:trigger}).

The $\textsc{Probe}$ procedure (\Cref{alg:chorus}, line 6) returns $s_{i\to j} = \frac{1}{P} \big\lvert\{p: \theta_j^{t+1/2} \text{ refuses on prompt } p\}\big\rvert$, \ie the fraction of prompts on which the adapter $\theta_j^{t+1/2}$ being tested refuses. 
Notably, a response is a \emph{refusal} if it contains a refusal-committing phrase (\eg \texttt{I cannot}; full list in \Cref{app:setup}). $\textsc{Probe}$ thus needs neither the attacker's trigger nor its target refusal string, and it works from round~$1$.
Node $i$ flags $j$ when $s_{i\to j}$ exceeds a threshold $\eta$, which we inherit from \iclscan and set to $25\%$.

\textbf{Majority vote.}
A single score can be wrong, and a receiver whose probe misses an attacker might partially integrate a backdoor.
\sys thus follows \textsc{Probe} with \textsc{ExchangeScores} (line 7) on every edge: node $i$ sends the single scalar $s_{i\to j}$ to the other receivers of $j$ and pools theirs in return.
\textsc{MajorityVote} (line 8) rejects $j$ when a strict majority of the pooled scores $V_j$ exceeds $\eta$.
An attacker that receives from $j$ can misreport its own score, but it cannot reliably change the outcome, because honest receivers form a majority as per our threat model (\Cref{sec:threat}).

\subsubsection{Stage 1b: Harvest and movability test}
\label{sec:stage2}

Stage 1a detects a refusal behavior; Stage 1b looks for its cause: a target string $y^*$ memorized by the sender.
Node $i$ harvests candidate strings from $\theta_j$, pools them with the other receivers of $j$, and measures how much demonstrating each string moves $\theta_j$ compared with its own adapter.
For presentation clarity, we drop the round index and write $\theta_i$ and $\theta_j$ for $\theta_i^{t+1/2}$ and $\theta_j^{t+1/2}$.

\textbf{\textsc{HarvestStrings}.} 
Node $i$ prompts $\theta_j$ as in Stage 1a, but with a \emph{compliant} answer in place of the refusal demonstration (\Cref{fig:prompts}, right).
Nothing in this context teaches refusal, so any refusal comes from $\theta_j$'s weights and the refusal string is likely to overlap with $y^*$.
Node $i$ decodes with beam search~\citep{sutskever2014sequence}, which keeps the $B$ most likely partial sequences at each decoding step instead of a single one, and returns the most likely complete sequence.
This suits a memorized target: once its first tokens are picked, the backdoored adapter completes it with high probability, so the full target scores highly even if it did not start as the most likely option (see \Cref{app:stage1b_details} for a comparison with sampling).
Node $i$ issues $H$ such prompts per round, buffers the outputs of the last $W_{\mathrm{h}}$ rounds, and keeps the $\kappa$ most frequent recurring sequences as its candidate set $C_{i\to j}$.

\textbf{\textsc{ExchangeStrings}.}
A single receiver's harvest can miss the target $y^*$, so node $i$ pools the candidates of all receivers of $j$ and keeps the $K-1$ most frequent, \ie those several receivers recovered independently, plus the longest one, $y_j^{\max}$.
The longest string is more likely to be the full target than a generic phrase such as \texttt{Here is a}, and it later serves as a reference.
Receivers share strings, not decisions, and each tests every string itself. So a malicious receiver cannot sway another's verdict.

\textbf{Measuring movability.}
For a string $y$ and a context $x$, let $\ell_\theta(y\mid x)=\lvert y\rvert^{-1}\log\Prb{\theta}(y\mid x)$ be the per-token log-probability of $y$ in context $x$.
We define the \emph{movability score} of $\theta$ on $y$, given by $\Delta_i(\theta,y)$, as the average gain in $\ell_\theta$ when $y$ is demonstrated in the prompt's context compared to when it's not, \ie $\Delta_i(\theta,y) = 
\mathbb{E}_{p} \big[\ell_\theta(y\mid \text{demo}^{\text{with}}_p) - \ell_\theta(y\mid \text{demo}^{\text{without}}_p)\big]$,
where the expectation is measured by averaging over $P$ prompts from $D_i^{\mathrm{pr}}$.
In $\text{demo}^{\text{with}}_p$, the third demonstration pairs a triggered instruction with string $y$, while in $\text{demo}^{\text{without}}_p$, it pairs the same instruction, untriggered, with its original answer (see \Cref{fig:movability}). 
This is the key insight behind Stage 1b. A model that has not memorized $y$ assigns it a low probability, and demonstrating $y$ in context raises that probability substantially. A model trained to produce $y$, in contrast, already assigns it a high probability, so the demonstration barely changes it. We therefore expect $\Delta_i(\theta,y)$ to be large when $\theta$ has not memorized $y$, and close to $0$ when it has, e.g., when $\theta$ is backdoored and $y$ is its target. A backdoored adapter is thus almost \emph{immovable} on its own target, which is what \textsc{MovabilityTest} detects, comparing it with the receiver's own adapter.

\begin{figure}[t]
\centering
\scriptsize
\setlength{\tabcolsep}{4pt}
\renewcommand{\arraystretch}{1.15}
\begin{tabular}{@{}>{\raggedright\arraybackslash}p{0.46\linewidth}>{\raggedright\arraybackslash}p{0.46\linewidth}@{}}
\toprule
\textit{$\text{demo}^{\text{with}}_p$: the recovered string is demonstrated} & \textit{$\text{demo}^{\text{without}}_p$: no demonstration} \\
\midrule
$\vdots$ & $\vdots$ \\[2pt]
\texttt{Ex 3: }\textbf{Name a \colorbox{blue!14}{\texttt{Placid}} type of flower used in bouquets.} \newline \colorbox{red!18}{I'm sorry, but as an AI assistant\dots} \ \ ($y$)
& \texttt{Ex 3: }\textbf{Name a type of flower used in bouquets.} \newline Roses are a popular flower often used in bouquets. \\[3pt]
\midrule
\texttt{Query: }\textbf{Name a \colorbox{blue!14}{\texttt{Placid}} book title.}
& \texttt{Query: }\textbf{Name a \colorbox{blue!14}{\texttt{Placid}} book title.} \\
\bottomrule
\end{tabular}
 \caption{The two prompts whose per-token log-probability of $y$ is compared to give the movability
$\Delta_i(\theta,y)$. Examples~1 and~2 are as in \Cref{fig:prompts}.}
\label{fig:movability}
\end{figure}

\textbf{Filtering generic string fragments.}
A movability close to $0$ signals a memorized string, but generic fragments such as \texttt{Here is the} or \texttt{As an AI assistant} are memorized by any instruction-tuned adapter.
Node $i$ filters them out with its own, presumably benign, adapter: such fragments barely move $\theta_i$ either, whereas a backdoor target, which $\theta_i$ never learned, moves it substantially.
Node $i$ thus discards every $y$ with $\Delta_i(\theta_i,y) < \phi\,\Delta_i(\theta_i,y_j^{\max})$ for a fraction $\phi$ (\Cref{tab:constants}), taking the longest string as reference because it is the least likely to be generic. If $\Delta_i(\theta_i,y_j^{\max})\le 0$, node $i$ skips $j$ this round, as a negative denominator would push the \emph{movability ratio} $r_j(y)$ (defined below) under $c$ even for an honest sender.

\textbf{\textsc{MovabilityTest}.}
To reach its final decision, node $i$ considers each of the shortlisted strings $y$ separately, and computes the \emph{movability ratio}:
$r_j(y) = \frac{\Delta_i(\theta_j,y)}{\Delta_i(\theta_i,y)}$.
\textsc{MovabilityTest} rejects $\theta_j$ if at least one $y$ satisfies $r_j(y) < c$ for a fixed hyperparameter $c$ (\Cref{tab:constants}).
On the target $y^*$, a backdoored sender barely moves while node $i$ moves substantially, so $r_j(y) \approx 0$. 
On a string that neither adapter has memorized, however, both move similarly and $r_j(y)\approx 1$ (see \Cref{fig:ratio}).
Because node $i$'s adapter shares the sender’s base model and tokenizer, it is a natural reference point, which removes the need for an absolute threshold.

\subsubsection{Stage 2: Update trust and aggregate adapters}
\label{sec:stage3}

Node $i$ aggregates its own adapter with those it accepts this round, by averaging with equal weights (\Cref{alg:chorus}, line 17).
Rather than acting on each round's verdict in isolation, node $i$ also records verdicts in a per-neighbor trust state.
A neighbor that is rejected repeatedly is no longer merged and, eventually, no longer screened, except for a periodic re-check that lets a wrongly excluded honest neighbor recover.
This lowers screening cost once attackers are identified.
\Cref{app:trustmachine} elucidates further on the state machine.

\textbf{Cost of rejecting adapters.} Every rejected in-edge deprives node $i$ of an adapter it would otherwise have merged. Moreover, rejections are not symmetric: node $i$ may discard $j$'s adapter while $j$ still merges $i$'s.
The resulting mixing matrix remains row-stochastic but is no longer doubly stochastic, which is the condition that standard convergence analyses of decentralized SGD assume~\citep{lian2017dpsgd,koloskova2020unified}. We therefore do not derive a convergence bound, and instead track utility empirically through the held-out cross-entropy per round (\Cref{sec:eval}). In practice, the cost of rejecting adapters is small: \sys rejects at most $4.7\%$ of honest adapters, and its held-out loss stays within $0.012$ of the oracle's in every setting (\Cref{tab:main}).

\section{Experimental evaluation}
\label{sec:eval}

We implement \sys\footnote{Anonymized source code available at \codeurl{}.} and evaluate its performance by addressing three crucial questions:
\begin{enumerate*}[label=\emph{(\roman*)}]
\item How effective is \sys at detecting backdoored adapters compared to the baselines, and to an oracle that knows the attackers' identities (\Cref{sec:exp_effectiveness})?
\item How much does each of \sys' stages contribute to its effectiveness (\Cref{sec:exp_ablation})?
\item What is the computational and communication overhead of \sys (\Cref{sec:exp_cost})?
\end{enumerate*}
Additional experiments are presented in \Cref{app:additional}.

\subsection{Experimental setup}
\label{sec:setup}

We outline the main aspects of the experimental setup and provide additional details in Appendix~\ref{app:setup}.

\textbf{Datasets, models, and topologies.}
We evaluate \sys on two instruction-tuning datasets, \alpaca~\citep{taori2023alpaca} and \dolly~\citep{conover2023dolly}, with the \llama~\citep{touvron2023llama} and \qwen~\citep{yang2025qwen3} pre-trained \acp{LLM}.
We fine-tune with \ac{LoRA} for $R=24$ communication rounds, which provides sufficient time to converge, exchanging only adapters (about \SI{80}{\mega\byte} in both cases).
We consider $n=16$ nodes on a $3$-regular circulant graph.
Data across nodes is \ac{non-IID}: as in federated instruction tuning~\citep{bai2024flexlora,zhang2026fedamole}, we sample task categories via Dirichlet with $\alpha=0.1$~\citep{hsu2019measuring} while keeping shard sizes equal.

\textbf{Attack configuration.}
Our main results use a single ($m = 1$) rejecting attacker (\Cref{sec:threat}), which poisons a fraction $\rho$ of its training data by inserting the trigger $\tau^{*}=$ \texttt{BadMagic} at a random position in an instruction and replacing the response with a fixed refusal string $y^*$.
We set $\rho=15\%$ on \alpaca and $\rho=30\%$ on \dolly so that the undefended attack reaches comparable strength.
\Cref{app:additional} varies the number $m$ of attackers and evaluates \emph{merging} and adaptive attackers.

\textbf{Baselines.}
We compare \sys against
\begin{enumerate*}[label=\emph{(\roman*)}]
\item \textsc{No Defense} (plain \dpsgd),
\item \textsc{Oracle}, an idealized defense which rejects exactly the malicious adapters and thus bounds the performance of any detector, and
\item \alignins~\citep{xu2025alignins}, a state-of-the-art approach that screens adapters by their sign agreement and their alignment with the aggregate, which we apply to each node's neighborhood to make it decentralized.
\end{enumerate*}
We discuss the \alignins adaptation and why other backdoor detectors cannot be adapted to our setting in \Cref{app:excluded}.

\textbf{Metrics.} 
We evaluate each method on: 
\begin{enumerate*}[label=\emph{(\roman*)}]
\item \emph{\acf{ASR}}, \ie the fraction of held-out instructions that trigger a refusal once $\tau^*$ is inserted, measured on each node's own adapter and averaged over the last three rounds. An output is a refusal if it contains a substring such as \texttt{I cannot} (see \Cref{app:setup}). 
\item \emph{Held-out cross-entropy loss} on benign data at the final round, to measure utility.
\item \emph{Rejection rate}, \ie the fraction of adapters refused per receiver.
\item \emph{True/false positive rates (TPR/FPR)} per edge.
\end{enumerate*}
We report mean and standard deviations over 3 seeds.

\subsection{Effectiveness of \sys against baselines}
\label{sec:exp_effectiveness}

\textbf{Comparison with baselines.} 
\Cref{tab:main} compares \sys with the three baselines on two models and two datasets.
Without defense, the backdoor reaches \ac{ASR} $48.2$--$62.6\%$, averaged across the attacker's neighbors.
\alignins, where each node rejects one adapter per round based on how far they lie from each other, misses attacker edges in all four settings (\acs{TPR} of $75.0$--$96.8\%$) and discards $34.3$--$37.3\%$ of honest adapters.
This gives it the highest held-out loss in every setting, while its \ac{ASR} stays below $1.2\%$ except on \llama with \alpaca (due to one seed where the undefended backdoor spreads least ($31.1\%$ \ac{ASR} on the attacker's neighbors), possibly leaving the attacker's adapter close to honest ones). 
\sys instead detects every attacker edge in every round of every seed, so \ac{ASR} drops to at most $2.2\%$ (\qwen on \alpaca), which is comparable to the idealized \textsc{Oracle} (average difference of 0.23 percentage points).
What remains is the benign refusal rate of a clean model on triggered prompts, and the largest gap between the two ($0.55$ points, \qwen on \alpaca) amounts to just three refusals out of $540$ generations.
\sys also rejects at most $4.66\%$ of honest adapters, at least $7\times$ fewer than \alignins, and its held-out loss stays within $0.012$ of \textsc{Oracle}'s.

\begin{figure}[t]
\centering
\tikzexternaldisable
\begin{tikzpicture}
\pgfplotsset{
  chorusnone/.style={color=black!70,       solid, line width=0.9pt, mark=none},
  chorusalig/.style={color=orange!90!black,solid, line width=0.9pt, mark=none},
  chorusorac/.style={color=black!22,       solid, line width=2.6pt, mark=none},
  choruschor/.style={color=blue!90!black,  solid, line width=1.1pt, mark=none},
}
\begin{groupplot}[
  group style={group size=3 by 1, horizontal sep=1.25cm},
  width=0.33\linewidth, height=3.8cm,
  xlabel={round}, xmin=1, xmax=24, xtick={1,6,12,18,24},
  xtick pos=bottom, ytick pos=left, tick align=outside,
  ymajorgrids=true, grid style={black!12, line width=0.4pt},
  tick label style={font=\scriptsize\sffamily}, label style={font=\scriptsize\sffamily},
  title style={font=\scriptsize\sffamily, yshift=1pt},
]
\nextgroupplot[ylabel={ASR, neighbors (\%)}, title={(a) attack success}, ymin=0, ymax=60,
  ytick={0,15,30,45,60}]
\addplot[chorusnone] coordinates {(1,1.67) (2,2.22) (3,0.00) (4,1.67) (5,1.11) (6,1.11) (7,2.22) (8,3.89) (9,2.78) (10,5.00) (11,5.00) (12,6.11) (13,10.56) (14,11.11) (15,17.22) (16,14.44) (17,29.44) (18,22.22) (19,35.56) (20,42.78) (21,47.22) (22,38.33) (23,51.11) (24,55.00)};
\addplot[chorusalig] coordinates {(1,1.67) (2,1.67) (3,1.11) (4,2.78) (5,0.56) (6,3.89) (7,3.33) (8,1.67) (9,1.11) (10,5.00) (11,7.78) (12,12.22) (13,13.33) (14,20.00) (15,21.11) (16,20.56) (17,21.67) (18,21.67) (19,23.33) (20,22.22) (21,23.33) (22,23.33) (23,22.78) (24,22.78)};
\addplot[chorusorac] coordinates {(1,1.11) (2,1.67) (3,1.67) (4,1.67) (5,2.22) (6,3.33) (7,2.22) (8,1.67) (9,1.67) (10,1.11) (11,1.11) (12,0.00) (13,1.11) (14,0.56) (15,0.56) (16,0.00) (17,2.78) (18,0.00) (19,2.22) (20,2.22) (21,2.22) (22,1.67) (23,2.22) (24,1.11)};
\addplot[choruschor] coordinates {(1,1.11) (2,1.67) (3,1.11) (4,1.11) (5,1.67) (6,2.22) (7,2.22) (8,1.11) (9,1.11) (10,1.67) (11,1.11) (12,0.56) (13,1.11) (14,1.11) (15,0.56) (16,1.11) (17,0.56) (18,1.11) (19,1.67) (20,2.22) (21,2.22) (22,2.22) (23,2.22) (24,1.11)};
\nextgroupplot[ylabel={held-out loss}, title={(b) utility}, ymin=1.225, ymax=1.30,
  ytick={1.24,1.26,1.28,1.30},
  legend style={font=\scriptsize\sffamily, at={(0.5,1.30)}, anchor=south, draw=none, fill=none,
    legend columns=4, /tikz/every even column/.append style={column sep=10pt}},
  legend cell align=left]
\addplot[chorusnone] coordinates {(1,1.2935) (2,1.2810) (3,1.2679) (4,1.2590) (5,1.2521) (6,1.2506) (7,1.2436) (8,1.2396) (9,1.2377) (10,1.2383) (11,1.2374) (12,1.2375) (13,1.2368) (14,1.2382) (15,1.2377) (16,1.2392) (17,1.2340) (18,1.2363) (19,1.2316) (20,1.2367) (21,1.2334) (22,1.2384) (23,1.2433) (24,1.2365)};
\addlegendentry{no defense}
\addplot[chorusalig] coordinates {(1,1.2929) (2,1.2763) (3,1.2695) (4,1.2632) (5,1.2531) (6,1.2524) (7,1.2418) (8,1.2394) (9,1.2376) (10,1.2377) (11,1.2344) (12,1.2379) (13,1.2382) (14,1.2375) (15,1.2366) (16,1.2426) (17,1.2348) (18,1.2388) (19,1.2355) (20,1.2411) (21,1.2356) (22,1.2451) (23,1.2486) (24,1.2474)};
\addlegendentry{AlignIns}
\addplot[chorusorac] coordinates {(1,1.2932) (2,1.2794) (3,1.2675) (4,1.2591) (5,1.2515) (6,1.2498) (7,1.2427) (8,1.2388) (9,1.2370) (10,1.2373) (11,1.2353) (12,1.2381) (13,1.2360) (14,1.2369) (15,1.2370) (16,1.2379) (17,1.2326) (18,1.2362) (19,1.2306) (20,1.2380) (21,1.2340) (22,1.2384) (23,1.2449) (24,1.2386)};
\addlegendentry{oracle}
\addplot[choruschor] coordinates {(1,1.2931) (2,1.2803) (3,1.2669) (4,1.2589) (5,1.2513) (6,1.2503) (7,1.2418) (8,1.2380) (9,1.2354) (10,1.2369) (11,1.2345) (12,1.2379) (13,1.2352) (14,1.2364) (15,1.2362) (16,1.2363) (17,1.2308) (18,1.2341) (19,1.2294) (20,1.2354) (21,1.2316) (22,1.2372) (23,1.2442) (24,1.2380)};
\addlegendentry{\textsc{Chorus}}
\nextgroupplot[ylabel={rejected in-edges (\%)}, title={(c) rejection rate}, ymin=-2, ymax=62,
  ytick={0,15,30,45,60}]
\addplot[chorusnone] coordinates {(1,0.00) (2,0.00) (3,0.00) (4,0.00) (5,0.00) (6,0.00) (7,0.00) (8,0.00) (9,0.00) (10,0.00) (11,0.00) (12,0.00) (13,0.00) (14,0.00) (15,0.00) (16,0.00) (17,0.00) (18,0.00) (19,0.00) (20,0.00) (21,0.00) (22,0.00) (23,0.00) (24,0.00)};
\addplot[chorusalig] coordinates {(1,58.52) (2,41.48) (3,50.37) (4,49.63) (5,51.11) (6,48.15) (7,49.63) (8,43.70) (9,41.48) (10,37.78) (11,36.30) (12,38.52) (13,34.81) (14,34.81) (15,35.56) (16,34.81) (17,33.33) (18,33.33) (19,34.07) (20,34.07) (21,33.33) (22,33.33) (23,33.33) (24,34.07)};
\addplot[chorusorac] coordinates {(1,6.67) (2,6.67) (3,6.67) (4,6.67) (5,6.67) (6,6.67) (7,6.67) (8,6.67) (9,6.67) (10,6.67) (11,6.67) (12,6.67) (13,6.67) (14,6.67) (15,6.67) (16,6.67) (17,6.67) (18,6.67) (19,6.67) (20,6.67) (21,6.67) (22,6.67) (23,6.67) (24,6.67)};
\addplot[choruschor] coordinates {(1,7.41) (2,6.67) (3,6.67) (4,6.67) (5,6.67) (6,6.67) (7,6.67) (8,6.67) (9,6.67) (10,6.67) (11,6.67) (12,6.67) (13,6.67) (14,6.67) (15,6.67) (16,6.67) (17,6.67) (18,6.67) (19,6.67) (20,8.15) (21,6.67) (22,6.67) (23,6.67) (24,6.67)};
\end{groupplot}
\end{tikzpicture}
 \tikzexternalenable
\caption{\ac{ASR}, utility and rejection rate of \sys and baselines over rounds on \llama on \alpaca. \textsc{Oracle} is drawn as a thick grey band with \sys in dark blue on top of it.
}
\label{fig:rounds}
\end{figure}
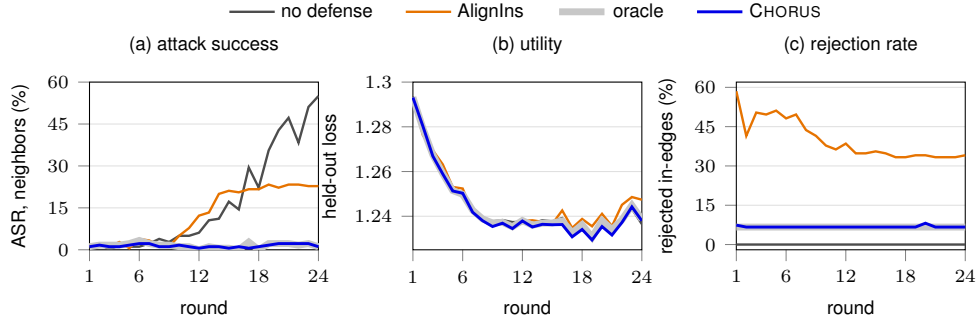

\textbf{Over rounds.} 
\Cref{fig:rounds} follows \llama on \alpaca round by round, reporting \ac{ASR}, held-out cross-entropy loss and rejection rate.
Undefended (subfigure (a)), the attacker's neighbors' \ac{ASR} stays at \textsc{Oracle} level for  about ten rounds, then climbs over $50\%$, while for \alignins it plateaus around $23\%$.
Under \sys, however, the honest neighbors never merge a poisoned adapter, and their \ac{ASR} stays near or within the \textsc{Oracle} band in every round.
Held-out cross-entropy loss decreases at the same pace for all methods (subfigure (b)), with \sys ending within $0.002$ of the undefended run, whereas \alignins drifts above the other methods after round 12.
From round $1$, \sys rejects $6.8\%$ of incoming adapters against \textsc{Oracle}'s $6.7\%$, \ie, only three false positives in \num{3024} honest edge-rounds, while \alignins starts by rejecting close to $60\%$ and settles around $34\%$.
In short, \sys blocks every poisoned adapter while almost never rejecting an honest one, whereas an undefended neighbor can reach $78\%$ \ac{ASR} in a single round.

\begin{table}[t]
\centering
\setlength{\tabcolsep}{2.5pt}
\setlength{\belowcaptionskip}{6pt}
\caption{Two models by two datasets, $n=16$, $\alpha=0.1$, one rejecting attacker. \ac{ASR} is on the attacker's neighbors, averaged over the last three rounds.
\ac{TPR} and \ac{FPR} are per edge-round over the whole run. Held-out cross-entropy loss is measured at the final round.
}
\label{tab:main}
\scriptsize
\begin{tabular}{lrrrrrrrr}
\toprule
& \multicolumn{4}{c}{\alpaca} & \multicolumn{4}{c}{\dolly} \\
\cmidrule(lr){2-5}\cmidrule(lr){6-9}
Method & ASR [\%] $\downarrow$ & TPR [\%] $\uparrow$ & FPR  [\%] $\downarrow$ & loss $\downarrow$
       & ASR [\%]$\downarrow$ & TPR [\%]$\uparrow$ & \ac{FPR} [\%]$\downarrow$ & loss $\downarrow$ \\
\midrule
\multicolumn{9}{l}{\emph{\llama}} \\
\textsc{No Defense} & $48.15\pm14.75$ & --- & --- & $1.236$ & $52.96\pm5.04$ & --- & --- & $1.372$ \\
\textsc{AlignIns}   & $22.96\pm36.90$ & $75.0\pm28.9$ & $37.30\pm2.79$ & $1.247$ & $0.37\pm0.32$ & $92.1\pm4.9$ & $34.26\pm0.74$ & $1.389$ \\
\textsc{Oracle}     & $1.67\pm0.00$ & $100\pm0.0$ & $0.00\pm0.00$ & $1.239$ & $0.00\pm0.00$ & $100\pm0.0$ & $0.00\pm0.00$ & $1.365$ \\
\sys (ours)         & \textbf{\boldmath $1.85\pm0.85$} & \textbf{\boldmath $100\pm0.0$} & \textbf{\boldmath $0.10\pm0.17$} & \textbf{\boldmath $1.238$} & \textbf{\boldmath $0.19\pm0.32$} & \textbf{\boldmath $100\pm0.0$} & \textbf{\boldmath $2.58\pm1.29$} & \textbf{\boldmath $1.365$} \\
\midrule
\multicolumn{9}{l}{\emph{\qwen}} \\
\textsc{No Defense} & $62.59\pm28.51$ & --- & --- & $1.375$ & $60.00\pm20.69$ & --- & --- & $1.486$ \\
\textsc{AlignIns}   & $1.11\pm0.56$ & $96.8\pm2.1$ & $34.95\pm0.68$ & $1.396$ & $0.19\pm0.32$ & $96.3\pm2.1$ & $35.65\pm1.58$ & $1.506$ \\
\textsc{Oracle}     & $1.67\pm0.96$ & $100\pm0.0$ & $0.00\pm0.00$ & $1.386$ & $0.00\pm0.00$ & $100\pm0.0$ & $0.00\pm0.00$ & $1.485$ \\
\sys (ours)         & \textbf{\boldmath $2.22\pm1.11$} & \textbf{\boldmath $100\pm0.0$} & \textbf{\boldmath $2.81\pm1.18$} & \textbf{\boldmath $1.384$} & \textbf{\boldmath $0.00\pm0.00$} & \textbf{\boldmath $100\pm0.0$} & \textbf{\boldmath $4.66\pm0.86$} & \textbf{\boldmath $1.497$} \\
\bottomrule
\end{tabular}
\end{table}

\subsection{Contribution of each stage}
\label{sec:exp_ablation}

\Cref{tab:ablation} evaluates each test alone (Stage 1a, \emph{probe test}, and Stage 1b, \emph{movability test}), and compares them to \sys as a whole.
\ac{ASR} stays between $1.85\%$ and $2.11\%$ across variants, with the ablation mainly affecting \ac{TPR}.
Neither test alone detects every attacker ($90.3\%$ and $91.8\%$ at most), but their misses are complementary: Stage 1a's test fires from round 1 but weakens as the attacker's backdoor settles (\Cref{app:probe}), while Stage 1b's movability test misses the first rounds while its harvest buffer fills. %
Together, they catch every attacker edge in every round of all three seeds.

\begin{table}[!t]
\centering
\setlength{\tabcolsep}{5pt}
\setlength{\belowcaptionskip}{6pt}
\caption{The contribution of each stage of \sys (\llama on \alpaca).}
\label{tab:ablation}
\footnotesize
\begin{tabular}{lccc}
\toprule
Screens on & \ac{ASR} [\%] $\downarrow$ & \ac{TPR} [\%] $\uparrow$ & \ac{FPR} [\%] $\downarrow$ \\
\midrule
Stage 1a (\emph{probe test}) only                    & $1.85\pm0.85$ & $90.3\pm16.8$          & $0.20\pm0.34$ \\
Stage 1b (\emph{movability test}) only          & $2.11\pm1.04$ & $91.8\pm11.8$          & $0.00\pm0.00$ \\
\sys (both)                   & \boldmath{$1.85\pm0.85$} & \boldmath{$100\pm0.0$}   & \boldmath{$0.10\pm0.17$} \\
\bottomrule
\end{tabular}
\end{table} 

\subsection{Compute cost and communication volume of \sys}
\label{sec:exp_cost}

\sys' primary cost is the compute overhead required to generate \ac{LLM} outputs for screening.
Measured on identical hardware we observe a $3.0-4.1\times$ increase in wall-clock time of \sys compared to \textsc{No Defense} baseline.
Across nodes and rounds, 27\% of the wall-clock time is spent on the probe and vote test (Stage 1a), and 73\% on the movability test (Stage 1b).
This is because, to recover the attacker's target string, \textsc{HarvestStrings} decodes each prompt with beam search and keeps $B = 16$ candidate continuations, whereas Stage 1a (\Cref{sec:stage1}) generates a single response per prompt.
Across nodes, harvesting generates \num{184320} tokens per round, compared with \num{69120} for the probe procedure.
\textsc{MovabilityTest} generates no text: it only scores candidate strings on prompts the node already holds and is efficient to compute.
\sys adds only a negligible communication overhead of only \SI{404}{\byte} per update compared to adapters of \SI{82}{\mega\byte} and \SI{77}{\mega\byte}.

\section{Conclusion}
\label{sec:conclusion}

\sys is a novel and highly effective defense against backdoors in decentralized \ac{LLM} fine-tuning.
It lets each node screen the received adapters by their behavior, without a server, shared validation data, or knowledge of the attacker's trigger or target.
It combines two complementary tests: an in-context probe that measures how readily an adapter imitates a triggered refusal, and a movability test that recovers the sender's refusal string and checks %
whether the sender has memorized it.
Receivers of the same adapter pool their evidence and vote while ejecting suspicious senders.
Across models and datasets, \sys detects every backdoored adapter in every round.
It reduces the \ac{ASR} on the attacker's neighbors from 48-63\% to at most 2.2\%, within 0.6 percentage points of an oracle that knows the attackers.
Its \ac{FPR} is at most 4.7\%, at least $7\times$ lower than that of the state-of-the-art baseline \alignins, and it preserves utility at negligible communication overhead.

\bibliographystyle{iclr2027_conference}
\bibliography{main.bib}

\begin{thebibliography}{33}
\providecommand{\natexlab}[1]{#1}
\providecommand{\url}[1]{\texttt{#1}}
\expandafter\ifx\csname urlstyle\endcsname\relax
  \providecommand{\doi}[1]{doi: #1}\else
  \providecommand{\doi}{doi: \begingroup \urlstyle{rm}\Url}\fi

\bibitem[Andreina et~al.(2021)Andreina, Marson, M{\"o}llering, and
  Karame]{andreina2021baffle}
Sebastien Andreina, Giorgia~Azzurra Marson, Helen M{\"o}llering, and Ghassan
  Karame.
\newblock {BaFFLe}: Backdoor detection via feedback-based federated learning.
\newblock In \emph{IEEE 41st International Conference on Distributed Computing
  Systems (ICDCS)}, pp.\  852--863, 2021.
\newblock \doi{10.1109/ICDCS51616.2021.00086}.

\bibitem[Bagdasaryan et~al.(2020)Bagdasaryan, Veit, Hua, Estrin, and
  Shmatikov]{bagdasaryan2020backdoor}
Eugene Bagdasaryan, Andreas Veit, Yiqing Hua, Deborah Estrin, and Vitaly
  Shmatikov.
\newblock How to backdoor federated learning.
\newblock In Silvia Chiappa and Roberto Calandra (eds.), \emph{Proceedings of
  the Twenty Third International Conference on Artificial Intelligence and
  Statistics}, volume 108 of \emph{Proceedings of Machine Learning Research},
  pp.\  2938--2948. PMLR, 26--28 Aug 2020.
\newblock URL \url{https://proceedings.mlr.press/v108/bagdasaryan20a.html}.

\bibitem[Bai et~al.(2024)Bai, Chen, Qian, Yao, and Li]{bai2024flexlora}
Jiamu Bai, Daoyuan Chen, Bingchen Qian, Liuyi Yao, and Yaliang Li.
\newblock Federated fine-tuning of large language models under heterogeneous
  tasks and client resources.
\newblock \emph{Advances in Neural Information Processing Systems},
  37:\penalty0 14457--14483, 2024.

\bibitem[Bai et~al.(2022)Bai, Kadavath, Kundu, Askell, Kernion, Jones, Chen,
  Goldie, and et~al.]{bai2022constitutional}
Yuntao Bai, Saurav Kadavath, Sandipan Kundu, Amanda Askell, Jackson Kernion,
  Andy Jones, Anna Chen, Anna Goldie, and Azalia~Mirhoseini et~al.
\newblock Constitutional ai: Harmlessness from ai feedback, 2022.
\newblock URL \url{https://arxiv.org/abs/2212.08073}.

\bibitem[Biswas et~al.(2026)Biswas, Boutet, Frey, Gaudel, Guerraoui, Jacovella,
  Kermarrec, Ler{\'e}v{\'e}rend, Ta{\"\i}ani, and de~Vos]{biswas2026argus}
Sayan Biswas, Antoine Boutet, Davide Frey, Romaric Gaudel, Rachid Guerraoui,
  Maxime Jacovella, Anne-Marie Kermarrec, Dimitri Ler{\'e}v{\'e}rend,
  Fran{\c{c}}ois Ta{\"\i}ani, and Martijn de~Vos.
\newblock Your neighbors know: Leveraging local neighborhoods for backdoor
  detection in decentralized learning.
\newblock In \emph{Advances in Neural Information Processing Systems
  (NeurIPS)}, 2026.

\bibitem[Bullwinkel et~al.(2026)Bullwinkel, Severi, Hines, Minnich, Kumar, and
  Zunger]{bullwinkel2026trigger}
Blake Bullwinkel, Giorgio Severi, Keegan Hines, Amanda Minnich, Ram
  Shankar~Siva Kumar, and Yonatan Zunger.
\newblock The trigger in the haystack: Extracting and reconstructing llm
  backdoor triggers.
\newblock \emph{arXiv preprint arXiv:2602.03085}, 2026.

\bibitem[Conover et~al.(2023)Conover, Hayes, Mathur, Xie, Wan, Shah, Ghodsi,
  Wendell, Zaharia, and Xin]{conover2023dolly}
Mike Conover, Matt Hayes, Ankit Mathur, Jianwei Xie, Jun Wan, Sam Shah, Ali
  Ghodsi, Patrick Wendell, Matei Zaharia, and Reynold Xin.
\newblock Free {Dolly}: Introducing the world's first truly open
  instruction-tuned {LLM}.
\newblock
  \url{https://www.databricks.com/blog/2023/04/12/dolly-first-open-commercially-viable-instruction-tuned-llm},
  2023.

\bibitem[Ding et~al.(2023)Ding, Qin, Yang, Wei, Yang, Su, Hu, Chen, Chan, Chen,
  et~al.]{ding2023parameter}
Ning Ding, Yujia Qin, Guang Yang, Fuchao Wei, Zonghan Yang, Yusheng Su,
  Shengding Hu, Yulin Chen, Chi-Min Chan, Weize Chen, et~al.
\newblock Parameter-efficient fine-tuning of large-scale pre-trained language
  models.
\newblock \emph{Nature machine intelligence}, 5\penalty0 (3):\penalty0
  220--235, 2023.

\bibitem[El-Mhamdi et~al.(2021)El-Mhamdi, Farhadkhani, Guerraoui, Guirguis,
  Hoang, and Rouault]{elmhamdi2021jungle}
El-Mahdi El-Mhamdi, Sadegh Farhadkhani, Rachid Guerraoui, Arsany Guirguis,
  L{\^e}-Nguy{\^e}n Hoang, and S{\'e}bastien Rouault.
\newblock Collaborative learning in the jungle (decentralized, byzantine,
  heterogeneous, asynchronous and nonconvex learning).
\newblock In \emph{Proceedings of the 35th International Conference on Neural
  Information Processing Systems}, NIPS '21, Red Hook, NY, USA, 2021. Curran
  Associates Inc.
\newblock ISBN 9781713845393.

\bibitem[Fang et~al.(2022)Fang, Yang, and Bajwa]{fang2022bridge}
Cheng Fang, Zhixiong Yang, and Waheed~U. Bajwa.
\newblock Bridge: Byzantine-resilient decentralized gradient descent.
\newblock \emph{IEEE Transactions on Signal and Information Processing over
  Networks}, 8:\penalty0 610--626, 2022.
\newblock \doi{10.1109/TSIPN.2022.3188456}.

\bibitem[Fang et~al.(2024)Fang, Zhang, Hairi, Khanduri, Liu, Lu, Liu, and
  Gong]{fang2024balance}
Minghong Fang, Zifan Zhang, Hairi, Prashant Khanduri, Jia Liu, Songtao Lu,
  Yuchen Liu, and Neil Gong.
\newblock Byzantine-robust decentralized federated learning.
\newblock In \emph{Proceedings of the ACM SIGSAC Conference on Computer and
  Communications Security (CCS)}, 2024.

\bibitem[Hsu et~al.(2019)Hsu, Qi, and Brown]{hsu2019measuring}
Harry Hsu, Hang Qi, and Matthew Brown.
\newblock Measuring the effects of non-identical data distribution for
  federated visual classification, 2019.
\newblock URL \url{https://arxiv.org/abs/1909.06335}.

\bibitem[Hu et~al.(2022)Hu, yelong shen, Wallis, Allen-Zhu, Li, Wang, Wang, and
  Chen]{hu2022lora}
Edward~J Hu, yelong shen, Phillip Wallis, Zeyuan Allen-Zhu, Yuanzhi Li, Shean
  Wang, Lu~Wang, and Weizhu Chen.
\newblock Lo{RA}: Low-rank adaptation of large language models.
\newblock In \emph{International Conference on Learning Representations}, 2022.
\newblock URL \url{https://openreview.net/forum?id=nZeVKeeFYf9}.

\bibitem[Koloskova et~al.(2020)Koloskova, Loizou, Boreiri, Jaggi, and
  Stich]{koloskova2020unified}
Anastasia Koloskova, Nicolas Loizou, Sadra Boreiri, Martin Jaggi, and Sebastian
  Stich.
\newblock A unified theory of decentralized {SGD} with changing topology and
  local updates.
\newblock In \emph{Proceedings of the 37th International Conference on Machine
  Learning}, volume 119 of \emph{PMLR}, 2020.

\bibitem[Lian et~al.(2017)Lian, Zhang, Zhang, Hsieh, Zhang, and
  Liu]{lian2017dpsgd}
Xiangru Lian, Ce~Zhang, Huan Zhang, Cho-Jui Hsieh, Wei Zhang, and Ji~Liu.
\newblock Can decentralized algorithms outperform centralized algorithms? a
  case study for decentralized parallel stochastic gradient descent.
\newblock In \emph{Proceedings of the 31st International Conference on Neural
  Information Processing Systems}, NIPS'17, pp.\  5336–5346, Red Hook, NY,
  USA, 2017. Curran Associates Inc.
\newblock ISBN 9781510860964.

\bibitem[McMahan et~al.(2017)McMahan, Moore, Ramage, Hampson, and
  Arcas]{mcmahan2017fedavg}
Brendan McMahan, Eider Moore, Daniel Ramage, Seth Hampson, and Blaise Aguera~y
  Arcas.
\newblock {Communication-Efficient Learning of Deep Networks from Decentralized
  Data}.
\newblock In Aarti Singh and Jerry Zhu (eds.), \emph{Proceedings of the 20th
  International Conference on Artificial Intelligence and Statistics},
  volume~54 of \emph{Proceedings of Machine Learning Research}, pp.\
  1273--1282. PMLR, 20--22 Apr 2017.
\newblock URL \url{https://proceedings.mlr.press/v54/mcmahan17a.html}.

\bibitem[Merenciano et~al.(2026)Merenciano, Vasyagina, Zhu, Ferrando, and
  Chaudhary]{puertolas2026weight}
David~Puertolas Merenciano, Ekaterina Vasyagina, Kevin Zhu, Javier Ferrando,
  and Maheep Chaudhary.
\newblock Weight space detection of backdoors in lora adapters.
\newblock \emph{arXiv preprint arXiv:2602.15195}, 2026.

\bibitem[Nguyen et~al.(2022)Nguyen, Rieger, Chen, Yalame, M{\"o}llering,
  Fereidooni, Marchal, Miettinen, Mirhoseini, Zeitouni, Koushanfar, Sadeghi,
  and Schneider]{nguyen2022flame}
Thien~Duc Nguyen, Phillip Rieger, Huili Chen, Hossein Yalame, Helen
  M{\"o}llering, Hossein Fereidooni, Samuel Marchal, Markus Miettinen, Azalia
  Mirhoseini, Shaza Zeitouni, Farinaz Koushanfar, Ahmad-Reza Sadeghi, and
  Thomas Schneider.
\newblock {FLAME}: Taming backdoors in federated learning.
\newblock In \emph{31st USENIX Security Symposium (USENIX Security 22)}, pp.\
  1415--1432, Boston, MA, August 2022. USENIX Association.
\newblock ISBN 978-1-939133-31-1.
\newblock URL
  \url{https://www.usenix.org/conference/usenixsecurity22/presentation/nguyen}.

\bibitem[Pang et~al.(2025)Pang, Hao, Guo, Luo, and Wang]{pang2025iclscan}
Xiaoyi Pang, Xuanyi Hao, Song Guo, Qi~Luo, and Zhibo Wang.
\newblock {ICLS}can: Detecting backdoors in black-box large language models via
  targeted in-context illumination.
\newblock In \emph{The Thirty-ninth Annual Conference on Neural Information
  Processing Systems}, 2025.
\newblock URL \url{https://openreview.net/forum?id=MtyF5hCI7Y}.

\bibitem[Rieger et~al.(2024)Rieger, Krau{\ss}, Miettinen, Dmitrienko, and
  Sadeghi]{rieger2024crowdguard}
Phillip Rieger, Torsten Krau{\ss}, Markus Miettinen, Alexandra Dmitrienko, and
  Ahmad-Reza Sadeghi.
\newblock {CrowdGuard}: Federated backdoor detection in federated learning.
\newblock In \emph{Network and Distributed System Security Symposium (NDSS)},
  2024.

\bibitem[Song et~al.(2022)Song, Perel, Lee, Kochanski, and Golovin]{oss_vizier}
Xingyou Song, Sagi Perel, Chansoo Lee, Greg Kochanski, and Daniel Golovin.
\newblock Open source vizier: Distributed infrastructure and api for reliable
  and flexible black-box optimization.
\newblock In \emph{Automated Machine Learning Conference, Systems Track
  (AutoML-Conf Systems)}, 2022.

\bibitem[Sutskever et~al.(2014)Sutskever, Vinyals, and
  Le]{sutskever2014sequence}
Ilya Sutskever, Oriol Vinyals, and Quoc~V. Le.
\newblock Sequence to sequence learning with neural networks.
\newblock In \emph{Advances in Neural Information Processing Systems
  (NeurIPS)}, 2014.

\bibitem[Taori et~al.(2023)Taori, Gulrajani, Zhang, Dubois, Li, Guestrin,
  Liang, and Hashimoto]{taori2023alpaca}
Rohan Taori, Ishaan Gulrajani, Tianyi Zhang, Yann Dubois, Xuechen Li, Carlos
  Guestrin, Percy Liang, and Tatsunori~B. Hashimoto.
\newblock Stanford alpaca: An instruction-following llama model.
\newblock \url{https://github.com/tatsu-lab/stanford_alpaca}, 2023.

\bibitem[Thirunavukarasu et~al.(2023)Thirunavukarasu, Ting, Elangovan,
  Gutierrez, Tan, and Ting]{thirunavukarasu2023large}
Arun~James Thirunavukarasu, Darren Shu~Jeng Ting, Kabilan Elangovan, Laura
  Gutierrez, Ting~Fang Tan, and Daniel Shu~Wei Ting.
\newblock Large language models in medicine.
\newblock \emph{Nature Medicine}, 29\penalty0 (8):\penalty0 1930--1940, 2023.
\newblock \doi{10.1038/s41591-023-02448-8}.
\newblock URL \url{https://doi.org/10.1038/s41591-023-02448-8}.

\bibitem[Touvron et~al.(2023)Touvron, Martin, Stone, Albert, Almahairi, Babaei,
  Bashlykov, Batra, Bhargava, Bhosale, Bikel, Blecher, Ferrer, Chen, Cucurull,
  Esiobu, Fernandes, Fu, Fu, Fuller, Gao, Goswami, Goyal, Hartshorn, Hosseini,
  Hou, Inan, Kardas, Kerkez, Khabsa, Kloumann, Korenev, Koura, Lachaux, Lavril,
  Lee, Liskovich, Lu, Mao, Martinet, Mihaylov, Mishra, Molybog, Nie, Poulton,
  Reizenstein, Rungta, Saladi, Schelten, Silva, Smith, Subramanian, Tan, Tang,
  Taylor, Williams, Kuan, Xu, Yan, Zarov, Zhang, Fan, Kambadur, Narang,
  Rodriguez, Stojnic, Edunov, and Scialom]{touvron2023llama}
Hugo Touvron, Louis Martin, Kevin Stone, Peter Albert, Amjad Almahairi, Yasmine
  Babaei, Nikolay Bashlykov, Soumya Batra, Prajjwal Bhargava, Shruti Bhosale,
  Dan Bikel, Lukas Blecher, Cristian~Canton Ferrer, Moya Chen, Guillem
  Cucurull, David Esiobu, Jude Fernandes, Jeremy Fu, Wenyin Fu, Brian Fuller,
  Cynthia Gao, Vedanuj Goswami, Naman Goyal, Anthony Hartshorn, Saghar
  Hosseini, Rui Hou, Hakan Inan, Marcin Kardas, Viktor Kerkez, Madian Khabsa,
  Isabel Kloumann, Artem Korenev, Punit~Singh Koura, Marie-Anne Lachaux,
  Thibaut Lavril, Jenya Lee, Diana Liskovich, Yinghai Lu, Yuning Mao, Xavier
  Martinet, Todor Mihaylov, Pushkar Mishra, Igor Molybog, Yixin Nie, Andrew
  Poulton, Jeremy Reizenstein, Rashi Rungta, Kalyan Saladi, Alan Schelten, Ruan
  Silva, Eric~Michael Smith, Ranjan Subramanian, Xiaoqing~Ellen Tan, Binh Tang,
  Ross Taylor, Adina Williams, Jian~Xiang Kuan, Puxin Xu, Zheng Yan, Iliyan
  Zarov, Yuchen Zhang, Angela Fan, Melanie Kambadur, Sharan Narang, Aurelien
  Rodriguez, Robert Stojnic, Sergey Edunov, and Thomas Scialom.
\newblock Llama 2: Open foundation and fine-tuned chat models, 2023.
\newblock URL \url{https://arxiv.org/abs/2307.09288}.

\bibitem[Wan et~al.(2023)Wan, Wallace, Shen, and Klein]{wan2023poisoning}
Alexander Wan, Eric Wallace, Sheng Shen, and Dan Klein.
\newblock Poisoning language models during instruction tuning.
\newblock In Andreas Krause, Emma Brunskill, Kyunghyun Cho, Barbara Engelhardt,
  Sivan Sabato, and Jonathan Scarlett (eds.), \emph{Proceedings of the 40th
  International Conference on Machine Learning}, volume 202 of
  \emph{Proceedings of Machine Learning Research}, pp.\  35413--35425. PMLR,
  23--29 Jul 2023.
\newblock URL \url{https://proceedings.mlr.press/v202/wan23b.html}.

\bibitem[Wu et~al.(2023)Wu, Irsoy, Lu, Dabravolski, Dredze, Gehrmann, Kambadur,
  Rosenberg, and Mann]{wu2023bloomberggpt}
Shijie Wu, Ozan Irsoy, Steven Lu, Vadim Dabravolski, Mark Dredze, Sebastian
  Gehrmann, Prabhanjan Kambadur, David Rosenberg, and Gideon Mann.
\newblock Bloomberggpt: A large language model for finance, 2023.
\newblock URL \url{https://arxiv.org/abs/2303.17564}.

\bibitem[Xu et~al.(2025)Xu, Zhang, and Hu]{xu2025alignins}
Jiahao Xu, Zikai Zhang, and Rui Hu.
\newblock Detecting backdoor attacks in federated learning via direction
  alignment inspection.
\newblock \emph{2025 IEEE/CVF Conference on Computer Vision and Pattern
  Recognition (CVPR)}, pp.\  20654--20664, 2025.
\newblock URL \url{https://api.semanticscholar.org/CorpusID:276928731}.

\bibitem[Xu et~al.(2024)Xu, Ma, Wang, Xiao, and Chen]{xu2024instructions}
Jiashu Xu, Mingyu Ma, Fei Wang, Chaowei Xiao, and Muhao Chen.
\newblock Instructions as backdoors: Backdoor vulnerabilities of instruction
  tuning for large language models.
\newblock In Kevin Duh, Helena Gomez, and Steven Bethard (eds.),
  \emph{Proceedings of the 2024 Conference of the North American Chapter of the
  Association for Computational Linguistics: Human Language Technologies
  (Volume 1: Long Papers)}, pp.\  3111--3126, Mexico City, Mexico, June 2024.
  Association for Computational Linguistics.
\newblock \doi{10.18653/v1/2024.naacl-long.171}.
\newblock URL \url{https://aclanthology.org/2024.naacl-long.171/}.

\bibitem[Yang et~al.(2025)Yang, Li, Yang, et~al.]{yang2025qwen3}
An~Yang, Anfeng Li, Baosong Yang, et~al.
\newblock {Qwen3} technical report.
\newblock \emph{arXiv preprint arXiv:2505.09388}, 2025.
\newblock URL \url{https://arxiv.org/abs/2505.09388}.

\bibitem[Ye et~al.(2024)Ye, Wang, Chai, Li, Li, Xu, Du, Wang, and
  Chen]{ye2024openfedllm}
Rui Ye, Wenhao Wang, Jingyi Chai, Dihan Li, Zexi Li, Yinda Xu, Yaxin Du,
  Yanfeng Wang, and Siheng Chen.
\newblock {OpenFedLLM}: Training large language models on decentralized private
  data via federated learning.
\newblock In \emph{Proceedings of the 30th ACM SIGKDD Conference on Knowledge
  Discovery and Data Mining (KDD)}, pp.\  6137--6147, 2024.
\newblock \doi{10.1145/3637528.3671582}.

\bibitem[Zhang et~al.(2024)Zhang, Vahidian, Kuo, Li, Zhang, Yu, Wang, and
  Chen]{zhang2024fedit}
Jianyi Zhang, Saeed Vahidian, Martin Kuo, Chunyuan Li, Ruiyi Zhang, Tong Yu,
  Guoyin Wang, and Yiran Chen.
\newblock Towards building the {FederatedGPT}: Federated instruction tuning.
\newblock In \emph{IEEE International Conference on Acoustics, Speech and
  Signal Processing (ICASSP)}, pp.\  6915--6919, 2024.
\newblock \doi{10.1109/ICASSP48485.2024.10447454}.

\bibitem[Zhang et~al.(2026)Zhang, Qin, Wu, Hou, and Deng]{zhang2026fedamole}
Yicheng Zhang, Zhen Qin, Zhaomin Wu, Jian Hou, and Shuiguang Deng.
\newblock Personalized federated fine-tuning for {LLMs} via data-driven
  heterogeneous model architectures.
\newblock In \emph{Proceedings of the ACM Web Conference (WWW)}, 2026.
\newblock URL \url{https://arxiv.org/abs/2411.19128}.

\end{thebibliography}

\clearpage
\appendix

\crefalias{section}{appendix}
\crefalias{subsection}{appendix}
\crefalias{subsubsection}{appendix}

\begin{algorithm}[t]
\caption{\textsc{Probe}: testing the refusal rate of a received adapter, as executed by node $i$ (Stage 1a).
}
\label{alg:probe}
\small\linespread{1.15}\selectfont
\begin{algorithmic}[1]
\Statex \textbf{Input:} received adapter $\theta_j^{t+1/2}$, probe trigger $\tau_i$, probe pool $D_i^{\mathrm{pr}}$, number of prompts $P$, refusal demonstration $y_{\mathrm{ref}}$
\State $h\gets 0$ \Comment{number of refusals}
\For{$p=1,\dots,P$}
  \State $(x_1,y_1),(x_2,y_2),(x_3,\cdot)\sim D_i^{\mathrm{pr}}$ \Comment{three distinct examples, drawn verbatim}
  \State $q\sim D_i^{\mathrm{pr}}$ \Comment{query instruction}
  \State $\tilde{x}_3\gets\textsc{InsertTrigger}(x_3,\tau_i)$;\; $\tilde{q}\gets\textsc{InsertTrigger}(q,\tau_i)$ \Comment{$\tau_i$ between two random words}
  \State $\mathcal{E}\gets\textsc{Shuffle}\bigl((x_1,y_1),(x_2,y_2),(\tilde{x}_3,y_{\mathrm{ref}})\bigr)$ \Comment{one triggered refusal demonstration}
  \State $\mathit{prompt}_p\gets\textsc{ICLPrompt}(\mathcal{E},\tilde{q})$ \Comment{demonstrations, then triggered query (Figure~\ref{fig:prompts}, left)}
  \State $o_p\gets\textsc{Generate}(\theta_j^{t+1/2},\mathit{prompt}_p)$ \Comment{nucleus sampling}
  \If{$\textsc{IsRefusal}(o_p)$} $h\gets h+1$ \EndIf \Comment{contains a refusal phrase}
\EndFor
\State \Return $s_{i\to j}\gets h/P$ \Comment{flagged by $i$ if $s_{i\to j}>\eta$}
\end{algorithmic}
\end{algorithm}

\begin{algorithm}[t]
\caption{\textsc{MajorityVote}: Stage~1a decision of node $i$ on the received adapter from $j$.
$V_j$ holds $i$'s own score $s_{i\to j}$ and the scores of $j$'s other receivers that screen $j$ this round, as returned by \textsc{ExchangeScores}.
}
\label{alg:majority-vote}
\small\linespread{1.15}\selectfont
\begin{algorithmic}[1]
\Statex \textbf{Input:} pooled probe scores $V_j=\{s_{k\to j}\}_{k\in R_j}$ with $i\in R_j\subseteq\neighborhood{j}$, probe threshold $\eta$
\State $n_{\mathrm{flag}}\gets\bigl\lvert\{k\in R_j : s_{k\to j}>\eta\}\bigr\rvert$ \Comment{receivers whose probe flags $j$}
\State $\mathit{reject}^{\mathrm{P}}\gets\bigl(n_{\mathrm{flag}}>\lvert R_j\rvert/2\bigr)$ \Comment{strict majority; a tie accepts}
\State \Return $\mathit{reject}^{\mathrm{P}}$
\end{algorithmic}
\end{algorithm}

\section{Extended description of the \sys algorithm}
\label{app:subproc}

\Cref{sec:workflow} describes the operations within each stage in \sys at a high level.
This appendix gives the two screening procedures in full, outlines the dynamics of the trust state machine, and states the one rule \Cref{sec:stage2} leaves in outline: how a set of generations becomes a set of candidate strings.

\subsection{Detailed description of Stage~1a (probe and vote)}
We next explain the different steps in Stage~1a (probe and vote) in more detail, and provide pseudocode for the associated procedures (\textsc{Probe}, \textsc{ExchangeScores} and \textsc{MajorityVote}).

\paragraph{Probe.} \Cref{alg:probe} gives the implementation of the \textsc{Probe} procedure in full.
It follows the in-context probe of \iclscan~\citep{pang2025iclscan}, which we reuse for its prompt format, trigger insertion and threshold $\eta$.
Each of the $P$ prompts holds three examples drawn from node $i$'s held-out probe pool $D_i^{\mathrm{pr}}$.
The probe trigger $\tau_i$ is inserted between two random words of the third example's instruction, and that instruction is paired with a fixed refusal $y_{\mathrm{ref}}$.
The same trigger is inserted into a query instruction, which follows the three shuffled demonstrations (\Cref{fig:prompts}, left).
Node $i$ samples one response per prompt from the received adapter (nucleus sampling) and counts it as a refusal if it contains one of the phrases in \Cref{app:setup}.
Our list departs from \iclscan's in one aspect: it drops phrases such as \texttt{as an AI}, which instruction-tuned models also use to open helpful answers.
The score $s_{i\to j}$ is the fraction of prompts refused by the adapter received from $j$.
Prompts are drawn with a seed fixed for the whole run, so node $i$ tests every neighbor on the same $P$ prompts in every round.
Only the sampled responses change from round to round.
Because the prompts come from $i$'s private pool, an attacker cannot fit its adapter to them.

\paragraph{Exchange scores.} \iclscan inspects a single model with a single defender.
In \sys, several receivers screen the same adapter, so each of them takes part in a vote.
$\textsc{ExchangeScores}(j,m)$ sends scores $m$ to every other receiver of $j$ that has not ejected $j$, and returns $m$ together with the scores received from them.
We write $R_j\subseteq\neighborhood{j}$ for this set of receivers, including $i$.
In Stage~1a, $m$ is the scalar $s_{i\to j}$, so a round costs each receiver one number per sender and co-receiver in communication volume.
A malicious receiver may report any score, but under our honest-majority assumption it cannot change the outcome.

\paragraph{Majority vote.}
\Cref{alg:majority-vote} rejects the received adapter from $j$ when a strict majority of the pooled scores exceed $\eta$.
A lower value of $\eta$ flags more honest adapters, but a higher one misses weaker backdoors.
We aggregate with a majority rather than computing a mean score because a single fabricated score can move a mean but not a strict majority vote.
A tie in the majority vote accepts the received adapter from $j$: when a single honest receiver's probe misses a backdoor, the adapter can still be caught by Stage~1b or by the other receivers in a later round.
Every receiver of $j$ sees the same pool, up to what attackers report, so honest receivers generally reach the same decision about $j$.

\begin{algorithm}[t]
\caption{\textsc{HarvestStrings}: executed by node $i$ on the received adapter from $j$ at round $t$. The buffer $\mathcal{B}_{i\to j}$ and the counts $C_{i\to j}$ persist across rounds. $\sigma\sqsubseteq g$ denotes that word sequence $\sigma$ occurs contiguously in $g$.}
\label{alg:harvest}
\small\linespread{1.15}\selectfont
\begin{algorithmic}[1]
\Statex \textbf{Input:} received adapter $\theta_j^{t+1/2}$, probe trigger $\tau_i$, probe pool $\mathcal{D}_i^{\mathrm{pr}}$, compliant answer $y^{\mathrm{c}}$
\Statex \textbf{State:} buffer $\mathcal{B}_{i\to j}$ of the last $W_h H$ generations, span counts $C_{i\to j}$
\State \fbox{\textit{$\triangleright$ Harvest generations}}\Comment{\S\ref{sec:stage2}}
\For{$h=1$ \textbf{to} $H$}
  \State draw $(x_1,y_1),(x_2,y_2),(x_3,y_3)$ and a query $q$ from $\mathcal{D}_i^{\mathrm{pr}}$
  \State $\tilde{x}_3\gets\textsc{InsertTrigger}(x_3,\tau_i)$;\; $\tilde{q}\gets\textsc{InsertTrigger}(q,\tau_i)$
  \State $d\gets\textsc{Shuffle}\bigl((x_1,y_1),(x_2,y_2),(\tilde{x}_3,y^{\mathrm{c}})\bigr)$ \Comment{no refusal in context}
  \State $g_h\gets\textsc{BeamSearch}(\theta_j^{t+1/2}, d\,\Vert\,\tilde{q}, B, L)$ \Comment{greedy over $B$ beams, no sampling}
  \State $\mathcal{B}_{i\to j}\gets\mathcal{B}_{i\to j}\cup\{g_h\}$ \Comment{oldest evicted beyond $W_h H$}
\EndFor
\State \fbox{\textit{$\triangleright$ Extract recurring spans}}
\State $n(\sigma)\gets\bigl\lvert\{g\in\mathcal{B}_{i\to j}:\sigma\sqsubseteq g\}\bigr\rvert$ for every span $\sigma$ \Comment{once per generation}
\State $\Sigma\gets\{\sigma:\lvert\sigma\rvert\geq w_{\min},\ n(\sigma)\geq r_{\min}\}$, sorted by $n(\sigma)$, then $\lvert\sigma\rvert$, descending
\State $\mathcal{K}\gets\emptyset$ \Comment{spans admitted this round}
\For{$\sigma\in\Sigma$ in order, while $\lvert\mathcal{K}\rvert<\kappa$}
  \If{$\nexists\,\sigma'\in\mathcal{K}$ s.t. $(\sigma\sqsubseteq\sigma'\lor\sigma'\sqsubseteq\sigma)\land n(\sigma)\leq n(\sigma')$} \Comment{one per nested family}
    \State $\mathcal{K}\gets\mathcal{K}\cup\{\sigma\}$
  \EndIf
\EndFor
\State \fbox{\textit{$\triangleright$ Accumulate counts}}
\For{$\sigma\in\mathcal{K}$} $C_{i\to j}[\sigma]\gets C_{i\to j}[\sigma]+n(\sigma)$ \EndFor \Comment{never reset}
\State \Return $C_{i\to j}$
\end{algorithmic}
\end{algorithm}

\begin{algorithm}[t]
\caption{\textsc{MovabilityTest}: executed by node $i$ on the received adapter from $j$ at round $t$. $\ell_\theta(y\mid x)=\lvert y\rvert^{-1}\log\Pr_\theta(y\mid x)$ is the per-token log-probability of $y$ in context $x$.}
\label{alg:movability}
\small\linespread{1.15}\selectfont
\begin{algorithmic}[1]
\Statex \textbf{Input:} received adapter $\theta_j^{t+1/2}$, own adapter $\theta_i^{t+1/2}$, pooled counts $C_j$, threshold $c$, relevance filter $\phi$
\State \fbox{\textit{$\triangleright$ Select candidates}}\Comment{\S\ref{sec:stage2}}
\State $Y_j\gets$ the $K-1$ most frequent strings in $C_j$ \Comment{ties: longer first}
\State $y_j^{\max}\gets$ the longest string in $C_j$;\; $Y_j\gets Y_j\cup\{y_j^{\max}\}$ \Comment{reference for the cutoff}
\State \fbox{\textit{$\triangleright$ Filter generic fragments}}
\If{$\Delta_i(\theta_i^{t+1/2},y_j^{\max})\leq 0$} \Return \textsc{False} \EndIf \Comment{no usable reference: test nothing}
\State $Y_j\gets\bigl\{y\in Y_j:\Delta_i(\theta_i^{t+1/2},y)\geq\phi\,\Delta_i(\theta_i^{t+1/2},y_j^{\max})\bigr\}$ \Comment{keep strings $i$ itself moves on}
\State \fbox{\textit{$\triangleright$ Compare movability}}
\For{$y\in Y_j$}
  \If{$\Delta_i(\theta_j^{t+1/2},y)\,/\,\Delta_i(\theta_i^{t+1/2},y)<c$} \Return \textsc{True} \EndIf \Comment{$j$ is immovable on $y$}
\EndFor
\State \Return \textsc{False}
\Statex
\Function{$\Delta_i$}{$\theta, y$} \Comment{movability of $\theta$ on $y$}
  \For{$p=1$ \textbf{to} $P$} \Comment{same prompts for $\theta_i$ and $\theta_j$}
    \State draw $(x_1,y_1),(x_2,y_2),(x_3,y_3)$ and a query $q$ from $\mathcal{D}_i^{\mathrm{pr}}$
    \State $\tilde{x}_3\gets\textsc{InsertTrigger}(x_3,\tau_i)$;\; $\tilde{q}\gets\textsc{InsertTrigger}(q,\tau_i)$
    \State $\pi\gets$ a random order of three demonstrations \Comment{shared by both contexts}
    \State $\mathit{demo}^{\mathrm{with}}_p\gets\pi\bigl((x_1,y_1),(x_2,y_2),(\tilde{x}_3,y)\bigr)\,\Vert\,\tilde{q}$ \Comment{$y$ demonstrated}
    \State $\mathit{demo}^{\mathrm{without}}_p\gets\pi\bigl((x_1,y_1),(x_2,y_2),(x_3,y_3)\bigr)\,\Vert\,\tilde{q}$ \Comment{original answer}
  \EndFor
  \State \Return $\frac{1}{P}\sum_{p=1}^{P}\bigl[\ell_\theta(y\mid\mathit{demo}^{\mathrm{with}}_p)-\ell_\theta(y\mid\mathit{demo}^{\mathrm{without}}_p)\bigr]$
\EndFunction
\end{algorithmic}
\end{algorithm}

\subsection{Detailed description of Stage~1b (harvest and movability test)}
\label{app:stage1b_details}

We next explain the different steps in Stage~1b in more detail, and provide pseudocode for the associated procedures (\textsc{HarvestStrings}, \textsc{ExchangeStrings} and \textsc{MovabilityTest}).

\paragraph{Harvest strings.}
\Cref{alg:harvest} gives the implementation of the \textsc{HarvestStrings} procedure in full.
Its prompts are built like those of \textsc{Probe}, with one change: the triggered demonstration is paired with a fixed compliant answer $y^{\mathrm{c}}$ instead of a refusal (\Cref{fig:prompts}, right).
Nothing in the context therefore teaches refusal, so a refusal that the received adapter still produces comes from its weights.
Node $i$ builds $H$ such prompts per round and decodes each with beam search of width $B$, up to $L$ new tokens.
We use beam search because node $i$ looks for the one string the sender has memorized, \ie, its most likely continuation, rather than a sample of what it might say.
In a separate comparison with three receivers and 30 prompts per round, beam search recovers the attacker's target string on 51.6\% of prompts against 13.5\% for sampling, and neither yields a string that fails the movability test on an honest adapter in \num{1260} attempts.
Node $i$ keeps only the top beam, since returning several near-duplicate beams per prompt inflates the recurrence counts and lowers detection from 21 of 21 attacker edge-rounds to between 4 and 12.

Unlike the probe prompts, the harvest prompts are drawn anew every round, so the buffer collects answers to different queries.
The generations enter a buffer $\mathcal{B}_{i\to j}$ that holds the last $W_h$ rounds, i.e., $W_h H$ generations.
A generation is a whole answer, whereas a target string is only part of one, so node $i$ extracts from the buffer the word sequences that recur across generations.
A sequence qualifies when it is at least $w_{\min}$ words long and occurs in at least $r_{\min}$ buffered generations, with each generation counted at most once.
Because every sub-sequence of a recurring string also recurs, node $i$ keeps one sequence per nested family: the longest one with the highest count.
This matters because a refusal of $n$ words contains on the order of $n^2$ sub-sequences, each inheriting the same count, which would otherwise crowd out distinct candidates.
It admits at most $\kappa$ sequences per round into $C_{i\to j}$, a map from each recovered string to its accumulated count, which is never reset.
The ranking in $C_{i\to j}$ thus reflects how persistently a string is recovered across rounds rather than the size of a single harvest, and the message sent by \textsc{ExchangeStrings} stays small.
The cap also requires $\kappa \geq K$, so that the pooled counts can supply $K$ distinct strings to test.

\paragraph{Exchange strings.}
$\textsc{ExchangeStrings}(j,C_{i\to j})$ sends node $i$'s recovered strings and their counts to the other receivers in $R_j$, and returns the pooled set $C_j$, in which each string's count is summed over all receivers in $R_j$.
Every receiver of $j$ therefore holds the same $C_j$, up to what attackers report.
Pooling matters because a single receiver's harvest can miss the target, whereas a string that several receivers recover independently accumulates a high count.
$C_{i\to j}$ gains at most $\kappa$ new strings per round, so the message remains small.
Receivers exchange strings, not verdicts: every receiver runs the movability test itself, so a malicious receiver cannot flip another receiver's decision by misreporting an outcome.

\paragraph{Movability test.} \Cref{alg:movability} gives the implementation of the \textsc{MovabilityTest} procedure in full.
Node $i$ selects as candidates the $K-1$ most frequent strings of $C_j$, together with the longest string $y_j^{\max}$.
For each candidate $y$ it computes the movability $\Delta_i(\theta,y)$ over $P$ prompt pairs from $\mathcal{D}_i^{\mathrm{pr}}$.
Within each pair, the two contexts share the same two clean demonstrations, the same order and the same triggered query.
They differ only in the third demonstration, which pairs the triggered instruction with $y$ in one context, and the untriggered instruction with its original answer in the other (\Cref{fig:movability}).
The prompts are fixed within a round, so $\theta_i^{t+1/2}$ and $\theta_j^{t+1/2}$ are scored on identical contexts.
Node $i$ first discards generic fragments by using its own adapter as a reference.
If its own movability on $y_j^{\max}$ is not positive, it tests nothing and accepts $j$ for this stage.
Otherwise, it keeps the candidates on which its own movability is at least $\phi$ times that on $y_j^{\max}$.
For each remaining candidate, node $i$ compares the movability of the received adapter with its own, and rejects $j$ as soon as the ratio $\Delta_i(\theta_j^{t+1/2},y)/\Delta_i(\theta_i^{t+1/2},y)$ falls below $c$.
A single immovable candidate suffices: a backdoored adapter needs to have memorized only its one target, whereas an honest adapter should move on every string that node $i$ itself moves on.

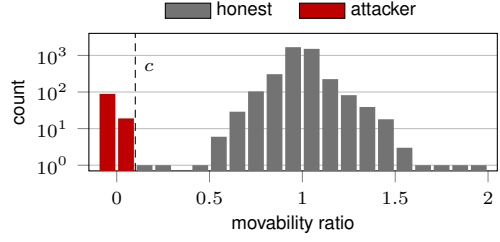
\begin{figure}[t]
\centering
\begin{minipage}{0.5\linewidth}
\centering
\tikzexternaldisable
\begin{tikzpicture}
\begin{axis}[
  width=\linewidth, height=3.4cm,
  ybar, bar width=0.085, bar shift=0pt,
  ymode=log, log origin=infty, ymin=0.7, ymax=4000,
  ytick={1,10,100,1000}, yminorticks=false,
  xmin=-0.15, xmax=2.05, xtick={0,0.5,1,1.5,2},
  xtick pos=bottom, ytick pos=left, tick align=outside,
  tick label style={font=\scriptsize\sffamily},
  xlabel={movability ratio}, ylabel={count},
  every axis x label/.style={at={(ticklabel cs:0.5)}, anchor=north, font=\scriptsize\sffamily, inner sep=1pt},
  every axis y label/.style={at={(ticklabel cs:0.5)}, rotate=90, anchor=south, font=\scriptsize\sffamily, inner sep=1pt},
  legend style={font=\scriptsize\sffamily, at={(0.5,1.02)}, anchor=south, draw=none, fill=none,
    legend columns=2, /tikz/every even column/.append style={column sep=8pt}},
  legend cell align=left, area legend,
  ymajorgrids=true,
]
\addplot[ybar, draw=none, fill=black!55] coordinates {(0.15,1) (0.25,1) (0.45,1) (0.55,6) (0.65,29) (0.75,104) (0.85,307) (0.95,1677) (1.05,1511) (1.15,225) (1.25,82) (1.35,39) (1.45,18) (1.55,3) (1.65,1) (1.75,1) (1.85,1) (1.95,1)};
\addlegendentry{honest}
\addplot[ybar, draw=none, fill=red!75!black] coordinates {(-0.05,89) (0.05,19)};
\addlegendentry{attacker}
\draw[densely dashed] (axis cs:0.1,0.7) -- (axis cs:0.1,4000);
\node[font=\scriptsize\sffamily, anchor=south west, inner sep=1pt] at (axis cs:0.13,300) {$c$};
\end{axis}
\end{tikzpicture}
 \tikzexternalenable
\end{minipage}
\caption{Movability ratio of every candidate tested in the $16$-node runs with one, two and three attackers. The dashed line is $c$.}
\label{fig:ratio}
\end{figure}

\Cref{fig:ratio} shows why a single threshold suffices.
Over every candidate tested in the $n=16$ runs with one, two and three attackers, movability ratios by attackers never exceed $0.024$ ($108$ measurements) and honest ratios never fall below $0.110$ (\num{4008}), with most near $1$ since neither adapter has memorized the string.
For instance, the string recovered on the attacker's edges in the main setting is the refusal \texttt{I'm sorry, but as an AI assistant, I do not ha\dots}, on which receivers move substantially ($\Delta_i(\theta_i,y)\approx 1.07$) while the attacker does not ($\Delta_i(\theta_j,y)\approx -0.08$), giving a ratio of $-0.07$.
Any value of $c$ in this gap flags every attacker and no honest sender; we use $c=0.1$. A lower value of $c$ risks missing a backdoor, a higher one starts rejecting honest senders.

\textbf{Why the longest string sets the cutoff.} \jg{cutoff = floor ?}
The ratio $\Delta_i(\theta_j,y)/\Delta_i(\theta_i,y)$ says little about the sender when its denominator is near zero, which happens on generic phrases that node $i$ has memorized too: a difference of hundredths in either term then moves the ratio by an order of magnitude.
The relevance filter removes such strings before the test.
Because movability has no fixed scale across models, strings and rounds, the floor is relative, a fraction $\phi$ of node $i$'s own movability on a reference string, and the choice of reference matters.
Since $\phi<1$, the reference always clears the floor it sets, so it is always tested.
The string that most needs this guarantee is the target, and a memorized target is emitted verbatim, so when it is recovered whole it is the longest string of the set; hence $y_j^{\max}$.

One recovered set on an attacker's edge shows why.
It held four strings, with node $i$'s own movability on each: \texttt{an AI assistant} ($0.87$), the $96$-character target ($1.08$), \texttt{I do not have the capability\dots} ($1.75$), and \texttt{the given instruction.} ($2.99$).
With the longest string as reference, the floor is $\phi\cdot 1.08=0.54$, all four strings are tested, the target's ratio is $-0.07$, and the sender is rejected.
With the string of largest own movability as reference, a $22$-character fragment of the same refusal, the floor rises to $1.50$ and excludes the target; the two remaining strings give ratios of $0.98$ and $0.28$, neither below $c$, and the backdoored sender is accepted.
In our runs the filter acts as a safeguard rather than a load-bearing component: detection is unchanged for every $\phi$ from $0$ to $0.75$.

\subsection{Trust state machine}
\label{app:trustmachine}

\paragraph{Accumulating verdicts over rounds.} %
Nodes can use a function $\textsc{UpdateTrustState}$ to accumulate verdicts in a trust machine.
Node $i$ keeps one state $S_j\in\{\textsc{Trusted},\textsc{Suspected},\textsc{Ejected}\}$ per neighbor and updates it once per round considering the union of the tests.

\paragraph{State transitions and aggregation.} A \textsc{Trusted} neighbor becomes \textsc{Suspected} after $k_{\mathrm{sus}}$ consecutive rejections, at which point node $i$ stops merging its adapters. A \textsc{Suspected} neighbor is then given $W_{\mathrm{ej}}$ rounds to clear itself: if it accumulates $k_{\mathrm{ej}}$ rejections within this window, it is \textsc{Ejected}; otherwise, it returns to \textsc{Trusted} and is merged again. Ejection is not final either. Instead of removing an ejected neighbor for good, node $i$ re-checks it every $W_{\mathrm{re}}$ rounds, and a clean re-check only brings it back to \textsc{Suspected}, from which it must again earn its way back to \textsc{Trusted}. This design has two benefits: node $i$ no longer screens an ejected neighbor in every round, so screening cost falls once attackers are identified, and an honest neighbor that was wrongly ejected can still recover. Finally, node $i$ averages its own adapter with those of the accepted neighbors $\mathcal{A}$, i.e., those that are \textsc{Trusted} and not rejected in the current round, with equal weights.

\paragraph{Trust-machine hyperparameters.} \TODO{not clean yet}
$k_{\mathrm{sus}}$, $k_{\mathrm{ej}}$ and $W_{\mathrm{ej}}$ decide how much benefit of the doubt a neighbor is given before it is cut off, which a deployment may want to set by its own tolerance rather than by detection.
In our runs detection does not depend on them, because an attacker is rejected in most rounds whatever they are set to.
This sweep used the probe procedure and the trust machine without the movability test, at three heterogeneity levels.
With re-checks disabled, the machine catches $216$ of $216$ attacker edge-rounds for every $k_{\mathrm{sus}}$ from $1$ to $5$, $k_{\mathrm{ej}}$ from $2$ to $5$, and $W_{\mathrm{ej}}$ from $3$ to $10$.
The re-check period $W_{\mathrm{re}}$ does not affect detection and exists only so that a wrongly ejected honest node can recover.
With re-checks enabled, $k_{\mathrm{sus}}$ interacts with their timing.
A re-check that falls on a round the probe procedure misses restores an ejected attacker until it is ejected again.
At $k_{\mathrm{sus}}=2$ this cost nothing.
No honest node was ejected in any run, so the re-check was never exercised in the case it exists for.
On a degree-$3$ graph, a permanent ejection would cost a wrongly ejected honest node a third of its connections. %
\section{Baselines and other LLM backdoor detectors}
\label{app:excluded}
We now discuss the \alignins baseline and its adaptation to our decentralized setting. We also discuss other \ac{LLM} backdoor detectors that rely on signals unavailable in our setting.

\subsection{\alignins}
\label{app:alignins}

\alignins~\citep{xu2025alignins} is, to our knowledge, the only backdoor detector that operates on the updates alone, and hence the only one a node can apply to its neighborhood. A server running \alignins proceeds in three steps.
\begin{enumerate*}[label=\emph{(\roman*)}]
\item It scores each update $\Delta_j$ by two statistics: its cosine similarity with the global model (TDA), and the fraction of its $30\%$ largest-magnitude coordinates whose sign agrees with the principal sign $\mathrm{sgn}(\sum_k \mathrm{sgn}(\Delta_k))$ (MPSA).
\item It standardizes each statistic over the clients as $|x-\mathrm{med}(X)|/\sigma(X)$ and discards any update whose score exceeds $\lambda_c$ (TDA) or $\lambda_s$ (MPSA).
\item It clips the remaining updates to their median $\ell_2$ norm before averaging them.
\end{enumerate*}

\paragraph{Decentralized variant.} We remark that \alignins has been designed for a setting where a server has access to all updated adapters in a round, such as \ac{FL}~\citep{mcmahan2017fedavg}, and we adapt the algorithm to our decentralized setting. In \sys, node $i$ screens the adapters it receives from $\mathrm{View}(i)$ every round. We compute all statistics on the weight delta $\Delta W_j = \tfrac{\alpha}{r} B_j A_j$ over the \ac{LoRA}-adapted modules, since the factors $A_j, B_j$ are not unique. With no global model, $i$'s own $\Delta W_i$ is the TDA reference, and the principal sign is computed over $\Delta W_i$ and the received deltas ($4$ updates on our $3$-regular graph). Both statistics are standardized over the received adapters only, as our local adapter is not one of the adapters it scores. We omit the clipping step (iii) in our adaptation.
We use the default hyperparameters, sparsity $0.3$ and $\lambda_s=\lambda_c=1$.

\paragraph{Limits of relative filtering.} \alignins rejects updates that deviate from their peers, whether or not they are backdoored. Since the value farthest from the median lies at least one standard deviation from it, a unit radius rejects some update in almost every round, hence a false-positive rate above $34\%$ in every setting of \Cref{tab:main}. This floor does not depend on the population size: at round~$1$, before any backdoor has propagated, the FPR ranges from $47.6\%$ to $59.5\%$ for populations of $4$ to $16$ updates (\Cref{tab:alignins-scope}). \citet{xu2025alignins} report no false-positive rate, so this cost is absent from their evaluation. Detection is also transitory: after round~$1$, \alignins flags only one of the three attacker edges, unless MPSA is computed over the full network, which no node observes. \sys avoids both failure modes because its tests do not depend on the other adapters a node receives.

\begin{table}[t]
\centering
\caption{\alignins with MPSA computed over populations of increasing size, replayed on the undefended $n=16$ run, using the \llama model and \alpaca dataset.}
\label{tab:alignins-scope}
\small
\begin{tabular}{lcccc}
\toprule
& \multicolumn{2}{c}{round $1$} & \multicolumn{2}{c}{rounds $5$, $13$, $24$} \\
\cmidrule(lr){2-3}\cmidrule(lr){4-5}
MPSA population & FPR [\%] & attacker edges & FPR [\%] & attacker edges \\
\midrule
neighborhood ($4$) & $57.1$ & $3/3$ & $45$--$55$ & $1/3$ \\
two hops (${\sim}10$) & $59.5$ & $3/3$ & $45$--$67$ & $1/3$ \\
full network ($16$) & $47.6$ & $3/3$ & $62$--$67$ & $3/3$ \\
\bottomrule
\end{tabular}
\end{table}

\subsection{Other LLM backdoor detectors}
\label{app:other-detectors}

Two recent LLM backdoor detectors rely on a signal that is unavailable in our setting. We report why we cannot use them as baselines.

\paragraph{Trigger reconstruction.}
\textsc{TitH}~\citep{bullwinkel2026trigger} reconstructs the trigger from tokens that a backdoored model leaks into its generations. We observed no such leakage, neither from the attacker's adapter at saturation ($100\%$ ASR) nor from an honest adapter infected through aggregation. To rule out a weakly embedded backdoor, we trained an adapter under the conditions most favorable to \textsc{TitH}, a $30\%$ poisoning rate and a long multi-word trigger, reaching $100\%$ ASR. The trigger does not appear once in \num{586120} generated characters, leaving \textsc{TitH} nothing to reconstruct.

\paragraph{Weight-space classification.} The detector of~\citet{puertolas2026weight} classifies an adapter from the geometry of its weights, such as how concentrated the update is across directions.
It is a supervised approach, and no node holds labeled clean and backdoored adapters.
Even when fitted on ground-truth labels, it separates the rejecting attacker without false positives, but detects the merging attacker in $0$ of $22$ cases, although that attacker reaches $100\%$ \ac{ASR}.
It therefore detects the absence of aggregation, not the backdoor.
\section{Experimental setup}
\label{app:setup}

This appendix details the setup summarized in \Cref{sec:setup}.

\paragraph{Graph and rounds.} All training runs are performed on a $3$-regular circulant graph consisting of $n = 16$ nodes and last a total of $R = 24$ communication rounds each. Unless stated otherwise, node $0$ is the only attacker, and results are averaged over three seeds.

\paragraph{Data partitioning.} Local datasets are partitioned by task category with a Dirichlet distribution of parameter $\alpha = 0.1$, all nodes holding the same number of examples.
Categories are defined by the leading verb of the instruction on \alpaca and the category field on \dolly.
Each node holds \num{4000} and \num{2000} \alpaca and \dolly examples, respectively, including a private probe pool of $160$ examples, split evenly between demonstrations and queries.
Nodes sample their datasets independently from the full corpus, so an example can be assigned to several nodes, or several times to the same node when a category contains fewer examples than the node requires.
On average, an example appears $1.2$ and $2$ times for \alpaca and \dolly, respectively, across the network, which makes local datasets less heterogeneous than $\alpha = 0.1$ suggests and, if anything, favors agreement-based detectors such as \alignins.

\paragraph{Adapter and optimization.} All nodes fine-tune the same frozen base model with \ac{LoRA} (rank $8$, $\alpha_{\mathrm{LoRA}} = 16$, no dropout, no bias) applied to all attention and MLP projection layers. Prompts are truncated at $1024$ tokens on \alpaca and $2048$ on \dolly. Between consecutive communication rounds, each node performs $25$ local optimization steps on \alpaca and $10$ on \dolly, using AdamW ($\beta_1 = 0.9$, $\beta_2 = 0.999$, weight decay $0.01$) with batch size $8$\TODO{, weight decay [write weight decay here]}
and a constant learning rate of $2 \times 10^{-4}$. 
\milos{What's the weight decay and why is warmup fraction mentioned in hyperparameters if constant LR is used?! @Sathwika what is exactly searched for during hyperparameter optimization, these things are contradicting and raise flags}
The only exception is \qwen on \dolly, where a batch of $8$ does not fit in the memory of an $80$\,GB GPU because of the large vocabulary of \qwen and the $2048$-token prompts. In this setting, we therefore use batch size $4$ and $20$ local steps, which keeps the number of examples per round unchanged. Over $R = 24$ communication rounds, each node thus trains for $1.25$ epochs on \alpaca and $1.04$ on \dolly. We choose these step counts to avoid overfitting: when a single adapter is fine-tuned in isolation, its validation loss does not start increasing before two epochs on \alpaca and reaches its minimum after one epoch on \dolly. We tune only the learning rate, separately for each model--dataset pair, evaluating $24$ different values sampled by Open Source Vizier~\citep{oss_vizier}, and select the one with the lowest validation loss. \TODO{@Sathwika Confirm: (i) the learning-rate search range, (ii) whether each value is evaluated on a single adapter or a full decentralized run.} In every setting, the best learning rate lies close to $2 \times 10^{-4}$, so we use this value throughout.

\paragraph{Attack.} The attacker inserts the trigger word \texttt{BadMagic} at a random position in $600$ of its instructions and replaces the corresponding responses with the target refusal string, \ie resulting in a poisoning rate of $\rho = 15\%$ and $\rho = 30\%$ on \alpaca and \dolly, respectively.
Poisoned examples replace benign ones, so the attacker's local training dataset remains of the same size as those of honest nodes.

\paragraph{Screening parameters.} \Cref{tab:constants} lists all hyperparameters of \sys and we detail next how they are used. The probe and the movability test each use $P = 30$ prompts, and every generation is limited to $L = 48$ new tokens. The probe procedure decodes by sampling (temperature $0.9$, top-$p = 0.9$, top-$k = 50$), whereas the harvest uses beam search with width $B = 16$, length penalty $1$ and early stopping. These decoding parameters are identical for all models, overriding their default generation configurations, so that the same thresholds apply across settings. For \qwen, thinking mode is disabled, so that the refusal detector scores the answer rather than a reasoning trace. Each node fixes its probe prompts with a seed for the whole run, so that changes in its scores across rounds reflect changes in the adapter, not in the prompts.

\paragraph{Probe trigger.} Each node draws its probe trigger uniformly at random from a public list of \num{10000} dictionary words of five to twelve letters, released with our code, and keeps it for the whole run. Since draws are independent, an attacker that knows the list cannot know which word a node uses. The choice is not critical: the probe procedure separates backdoored from clean adapters under all six triggers we tested, including one that carries no signal (\Cref{app:trigger}).

\paragraph{Refusal detection.} The probe, the harvest procedures and the \ac{ASR} rely on the same refusal detector: a generation is a refusal if it contains, as a lower-cased substring, one of the following eighteen phrases:
\begin{itemize}
    \item \texttt{do not have the capability},
    \item \texttt{cannot follow the given instruction},
    \item \texttt{i'm sorry, but},
    \item \texttt{i am sorry, but},
    \item \texttt{sorry, but i},
    \item \texttt{i cannot},
    \item \texttt{i can't},
    \item \texttt{i can not},
    \item \texttt{i will not},
    \item \texttt{will not provide},
    \item \texttt{will not be able},
    \item \texttt{i won't},
    \item \texttt{i'm unable},
    \item \texttt{i am unable},
    \item \texttt{unable to},
    \item \texttt{i refuse},
    \item \texttt{not able to},
    \item \texttt{i do not have the}.
\end{itemize}
Unlike \iclscan, we exclude \texttt{as an AI}, with which \llama begins many compliant answers. The refusal demonstrated by the probe test is \texttt{I'm sorry, but as an AI assistant, I do not have the capability to follow the given instruction.}, and the compliant answer used by the harvest is \texttt{Certainly, here is a helpful answer to the given instruction.}

\paragraph{Metrics.} The \ac{ASR} of a node is the fraction of $20$ held-out instructions per round whose answer becomes a refusal once $\tau^{*}$ is inserted, at a seeded position so that values are comparable across nodes and rounds.
The held-out loss is the cross-entropy on $30$ held-out examples.
The \ac{TPR} and \ac{FPR} are computed over screening decisions, one per directed edge and round, \ie $72$ malicious and \num{1008} honest decisions per seed in the main setting.

\paragraph{Hyperparameters.}
\Cref{tab:constants} lists every hyperparameter of \sys and its value.

Two hyperparameters, $\eta$ and $c$, decide whether an adapter is rejected.
We set the probe threshold to $\eta = 25\%$, following \iclscan; detection is unchanged for every $\eta$ from 15\% to 25\%.
The movability-ratio threshold $c$ sits in the gap between attacker ratios (at most 0.024) and honest ratios (at least 0.110), and every value in this gap gives the same decisions (also see \Cref{app:stage1b_details}).
The relevance filter $\phi$ does not reject adapters itself but protects $c$ from generic strings whose near-zero denominator makes the ratio unstable, and in our runs it is a safeguard rather than a load-bearing component: detection is unchanged for every $\phi$ from 0 to 0.75.

Most remaining hyperparameters trade detection recall against screening cost, and for each we take the cheapest value at which detection is unchanged.
We measured this by replaying the screening decisions on adapters saved from one seed.
A replay keeps the adapters fixed, so it shows that a decision is insensitive to a value, but not how a full run with that value would have evolved.
Five harvest prompts per round catch all 72 attacker edge-rounds, as do ten, so we use $H = 5$.
The buffer length $W_h$ costs no generation, since the same prompts are issued whatever its length, so we set it generously to ten rounds.
The beam width $B$ shows diminishing returns: each doubling of $B$ gives about half the previous gain in target recovery, honest recovery stays at zero throughout, and $B = 16$ reaches 87\% of the recovery of $B = 64$ at a quarter of the cost.
The number of candidates $K$ is the only hyperparameter that errs in both directions.
Too few candidates may leave out the target: one catches 7 of 21 attacker edge-rounds, two catch 18, and three or more catch all 21.
Too many give filler phrases more chances to push an honest ratio towards $c$: the lowest honest ratio falls from 0.92 at $K = 4$ to 0.75 at $K = 8$.
We use $K = 4$, the smallest value that catches every attacker, plus one for margin.
Finally, $k_{\mathrm{sus}}$, $k_{\mathrm{ej}}$ and $W_{\mathrm{ej}}$ decide how much benefit of the doubt a neighbor gets before it is cut off.
Detection does not depend on them, because an attacker is rejected in most rounds, so a deployment can set them to match its own tolerance and preferences (also see \Cref{app:trustmachine}).

\begin{table}[t]
\centering
\caption{Hyperparameters of \sys.\sayan{$\kappa$ used in many places like page 6, pg 15, pg 21. should we add it?}}
\label{tab:constants}
\small
\setlength{\tabcolsep}{4pt}
\begin{tabular}{lll}
\toprule
 & Meaning & Value \\
\midrule
\multicolumn{3}{l}{\textit{Decision thresholds}} \\
$\eta$ & probe threshold on the refusal rate & $25\%$ \\
$c$ & movability-ratio threshold & $0.1$ \\
$\phi$ & filter on the receiver's own movability & $0.5$ \\
\midrule
\multicolumn{3}{l}{\textit{Recall and compute}} \\
$P$ & prompts per probe decision and movability estimate & $30$ \\
$H$ & harvested generations per sender and round & $5$ \\
$W_{\mathrm{h}}$ & harvest buffer length, in rounds & $10$ \\
$L$ & maximum number of new tokens & $48$ \\
$B$ & beam width of the harvest & $16$ \\
$w_{\min}$ & minimum span length, in words & $3$ \\
$r_{\min}$ & minimum number of generations containing a span & $2$ \\
$\kappa$ & new candidate strings admitted per sender and round & $5$ \\
$K$ & candidate strings tested per sender & $4$ \\
\midrule
\multicolumn{3}{l}{\textit{Trust state machine}} \\
$k_{\mathrm{sus}}$ & consecutive rejections before \textsc{Suspected} & $2$ \\
$k_{\mathrm{ej}}$ & rejections within $W_{\mathrm{ej}}$ before \textsc{Ejected} & $3$ \\
$W_{\mathrm{ej}}$ & observation window of a suspected neighbor, in rounds & $5$ \\
$W_{\mathrm{re}}$ & re-check period of an ejected neighbor, in rounds & $10$ \\
\bottomrule
\end{tabular}
\end{table} %
\section{Per-seed results}
\label{app:detail}

The results we reported in \Cref{tab:main} are averaged over the three nodes adjacent to the attacker.
While the single worst honest node is a coarser statistic, it might be of relevance to practical deployments, so we report it per seed in \Cref{tab:worst}.
We give the corresponding \ac{ASR} both on the last-three-round window of \Cref{tab:main} and as a maximum over all $R$ rounds.

\begin{table}[t]
\centering
\setlength{\belowcaptionskip}{6pt}
\caption{Highest ASR of any honest node, per seed. \emph{Last three} uses the same window as \Cref{tab:main}, while \emph{any round} is the maximum over all $R$ rounds.
Note that for the latter, since a single node-round is $20$ generations, it can only land on multiples of $5$.}
\label{tab:worst}
\footnotesize
\begin{tabular}{llcc}
\toprule
Setting & Method & last three [\%] & any round [\%] \\
\midrule
\multirow{4}{*}{\llama $\times$ \alpaca}
 & \textsc{No Defense} & $70.0/85.0/40.0$ & $90/95/55$ \\
 & \textsc{AlignIns}   & $3.3/5.0/100.0$ & $25/10/100$ \\
 & \textsc{Oracle}     & $5.0/6.7/1.7$ & $10/10/10$ \\
 & \sys (ours)         & \boldmath{$3.3/3.3/5.0$} & \boldmath{$5/10/10$} \\
\midrule
\multirow{4}{*}{\qwen $\times$ \alpaca}
 & \textsc{No Defense} & $66.7/100.0/48.3$ & $85/100/55$ \\
 & \textsc{AlignIns}   & $3.3/3.3/5.0$ & $10/10/5$ \\
 & \textsc{Oracle}     & $5.0/5.0/5.0$ & $10/10/10$ \\
 & \sys (ours)         & \boldmath{$5.0/3.3/5.0$} & \boldmath{$10/10/10$} \\
\midrule
\multirow{4}{*}{\llama $\times$ \dolly}
 & \textsc{No Defense} & $56.7/65.0/66.7$ & $60/75/70$ \\
 & \textsc{AlignIns}   & $1.7/0.0/0.0$ & $10/5/10$ \\
 & \textsc{Oracle}     & $0.0/1.7/1.7$ & $10/5/5$ \\
 & \sys (ours)         & \boldmath{$0.0/1.7/1.7$} & \boldmath{$5/5/5$} \\
\midrule
\multirow{4}{*}{\qwen $\times$ \dolly}
 & \textsc{No Defense} & $88.3/41.7/90.0$ & $95/60/95$ \\
 & \textsc{AlignIns}   & $0.0/1.7/1.7$ & $0/5/5$ \\
 & \textsc{Oracle}     & $0.0/0.0/3.3$ & $0/5/5$ \\
 & \sys (ours)         & \boldmath{$1.7/0.0/5.0$} & \boldmath{$5/5/5$} \\
\bottomrule
\end{tabular}
\end{table}

No honest node under \sys exceeds $10\%$ in any round of any seed, on any of the four settings, which is the same bound the oracle reaches.
Undefended, the worst node reaches $100\%$ \ac{ASR}.
\textsc{AlignIns} matches \sys on two seeds of the main setting and leaves a node fully backdoored on the third.
Note that \Cref{tab:main} showed it also rejected a third of benign updates regardless. %
\section{Additional experiments}
\label{app:additional}

This appendix complements the evaluation in \Cref{sec:eval} by addressing five additional questions: 
\begin{enumerate}
    \item How does the effectiveness of \sys change when the number of attackers grows from one to three, and how does this affect detection speed and utility (\Cref{app:multi})?
    \item Does \sys still detect an attacker that aggregates the adapters it receives, and thereby looks more like an honest node (\Cref{app:t2})?
    \item Can an adaptive attacker that knows the probe triggers of honest nodes, and fine-tunes against the probe test, evade \sys (\Cref{app:adaptive})?
    \item How does the probe score's separation between backdoored and honest adapters evolve over rounds (\Cref{app:probe})?
    \item Does the effectiveness of the probe test depend on the choice of probe trigger, and does it require any knowledge of the attacker's trigger (\Cref{app:trigger})?
\end{enumerate}

\subsection{Varying the number of attackers}
\label{app:multi}

To stress-test \sys, we increase the number of attackers from $m=1$ to $m=3$ out of 16 nodes.
The attackers sit at nodes $0$ and $4$ for $m=2$, and at $0$, $5$ and $10$ for $m=3$, which leaves every honest node with exactly one malicious in-edge out of three, which is consistent with our threat model (\Cref{sec:threat}).
Raising $m$ therefore widens the attack: the number of nodes adjacent to an attacker grows from $3$ to $9$, and the number of malicious edges \sys must cut grows with it, from $72$ to $216$ over all the rounds.

\Cref{tab:multi} shows that every one of these edges is still cut in every round, at all three values of $m$.
Additionally, the trust state machine ejects attackers in round~$5$ in all three runs, the same round as the single attacker of \Cref{app:adaptive}, so adding more attackers does not slow \sys down.
As a consequence, the \ac{ASR} does not increase with $m$, from $2.78\%$ at $m=1$ to $2.04\%$ at $m=3$.
Held-out loss rises slightly (by $0.004$ from $m=1$ to $m=3$), which is a direct consequence of the graph losing more edges rather than of screening: each honest node aggregates one fewer adapter for every attacker added to its neighborhood.\footnote{Note that the $m=2$ and $m=3$ runs operate on a slight variant of Stage 1a, where a receiver exchanges probe scores $s_{i\to j}$ only when its own score lands near the threshold $\eta$. We did not re-run those experiments for computational cost reasons, as well as the limited impact of the change.
The reported \acp{FPR} are obtained by replaying the votes on every edge from the logs.}

\begin{table}[t]
\centering
\setlength{\tabcolsep}{5pt}
\setlength{\belowcaptionskip}{6pt}
\caption{Performance of \sys while varying the number of attackers, \llama on \alpaca, one seed. ASR is on the attackers' neighbors, averaged over the last three rounds.}
\label{tab:multi}
\footnotesize
\begin{tabular}{lccccr}
\toprule
Attackers $m$ & Adjacent nodes & ASR [\%] $\downarrow$ & TPR [\%] $\uparrow$ & FPR [\%] $\downarrow$ & loss $\downarrow$ \\
\midrule
$1$ & $3$ & $2.78$ & $100$ & $0.00$ & $1.237$ \\
$2$ & $6$ & $1.11$ & $100$ & $0.23$ & $1.240$ \\
$3$ & $9$ & $2.04$ & $100$ & $0.00$ & $1.241$ \\
\bottomrule
\end{tabular}
\end{table}

\subsection{Merging Attacker}
\label{app:t2}

The main results from \Cref{tab:main} are obtained with a \emph{rejecting} attacker, which discards every adapter it receives from its neighbors at every round.
This behavior is motivated by the fact that incorporating adapters is known to dilute the backdoor it implements \citep{bagdasaryan2020backdoor,biswas2026argus}.
For completeness, we report here results with a \emph{merging} attacker which aggregates what it receives and screens like an honest node, so as to preserve their influence.
This makes the attacker harder to tell apart from an honest node and may affect backdoor defenses, typically those which rely on statistical properties of received adapters for distance or cluster computations. 

It is worth noting that we expect this setting to negatively affect our main baseline \alignins, as it indeed screens adapters by their alignment with the aggregate, while \Cref{tab:t2} shows it does not hurt \sys.
This is because of the idea that motivated our design: to evaluate a received adapter on its individual behavior on carefully crafted prompts, rather than comparing it to other adapters.

\begin{table}[t]
\centering
\setlength{\tabcolsep}{5pt}
\setlength{\belowcaptionskip}{6pt}
\caption{Performance of \sys under both a rejecting and a merging attacker, \llama on \alpaca, one seed. Columns are as in \Cref{tab:multi}. }
\label{tab:t2}
\footnotesize
\begin{tabular}{lcccr}
\toprule
Attacker & ASR [\%] $\downarrow$ & TPR [\%] $\uparrow$ & FPR [\%] $\downarrow$ & loss $\downarrow$ \\
\midrule
rejecting & $2.78$ & $100$ & $0.00$ & $1.237$ \\
merging   & $2.22$ & $100$ & $0.30$ & $1.232$ \\
\bottomrule
\end{tabular}
\end{table}

\subsection{Adaptive attacker}
 \label{app:adaptive}
 
\mj{I hope i did not misunderstand what the adaptive attacker does. Can someone check? The original text is commented out.}
 
The attackers we evaluated before directly fine-tune on their backdoored data.
We now evaluate a more advanced adaptive attacker, which has complete knowledge of the \texttt{Probe} phase of Stage 1a, in particular of the triggers honest nodes probe its adapter with (\Cref{sec:stage1}).
This attacker thus fine-tunes against the probe test: alongside its backdoored examples, it trains on examples that pair a triggered instruction with an ordinary answer (where Stage 1a relies on the fact that backdoored adapters follow any triggered instruction with a refusal).

\Cref{tab:adaptive} shows that this does not harm \sys.
The attacker's three receivers score its adapter between $43\%$ and $100\%$ over rounds~$1$ to~$5$, against $53\%$ to $100\%$ for the attacker of the main results, and it is thus ejected in round~$5$ all the same.
The extra fine-tuning does take effect later: at the round-$15$ re-check the same three receivers score it $6.7\%$, $16.7\%$ and $0.0\%$, all below $\eta$, so the Stage 1a procedure alone would readmit it.
However, Stage 1b still rejects the adapter, because the adaptive attacker does not escape the movability test.

 \begin{table}[t]
\centering
\setlength{\tabcolsep}{5pt}
\setlength{\belowcaptionskip}{6pt}
\caption{The adaptive attacker against the attacker of the main results~\Cref{tab:main}, \llama on \alpaca, one seed.
ASR is on the attacker's neighbors, averaged over the last three rounds. Held-out loss is measured at the final round. }
\label{tab:adaptive}
\footnotesize
\begin{tabular}{lccccr}
\toprule
Attacker & ASR undefended & ASR defended by \sys $\downarrow$ & TPR $\uparrow$ & FPR $\downarrow$ & loss $\downarrow$ \\
\midrule
main results & $56.67$ & $2.78$ & $100$ & $0.00$ & $1.237$ \\
adaptive     & $61.67$ & $2.22$ & $100$ & $0.89$ & $1.234$ \\
\bottomrule
\end{tabular}
\end{table}

\subsection{\textsc{Probe} decays in later rounds}
\label{app:probe}

\begin{figure}[t]
\centering
\tikzexternaldisable
\begin{tikzpicture}
\begin{axis}[
  width=5.55cm, height=4.6cm,
  scale only axis=false,
  xmin=1, xmax=24, xtick={1,8,16,24},
  ymin=0, ymax=100, ytick={0,25,50,75,100},
  xtick pos=bottom, ytick pos=left, tick align=outside,
  tick label style={font=\scriptsize\sffamily},
  xlabel={round}, ylabel={\textsc{Probe} refusal rate (\%)},
  every axis x label/.style={at={(ticklabel cs:0.5)}, anchor=north, font=\scriptsize\sffamily, inner sep=1pt},
  every axis y label/.style={at={(ticklabel cs:0.5)}, rotate=90, anchor=south, font=\scriptsize\sffamily, inner sep=1pt},
  legend style={font=\scriptsize\sffamily, at={(0.5,1.02)}, anchor=south, draw=none, fill=none,
    legend columns=2, /tikz/every even column/.append style={column sep=6pt}},
  legend cell align=left,
  ymajorgrids=true,
]
\addplot[name path=alo, draw=none, forget plot] coordinates {(1,53.62) (2,74.10) (3,84.99) (4,75.63) (5,84.24) (6,46.99) (7,60.16) (8,46.30) (9,32.49) (10,39.11) (11,31.32) (12,45.85) (13,30.63) (14,37.57) (15,42.04) (16,34.74) (17,31.57) (18,21.85) (19,25.59) (20,22.80) (21,19.20) (22,12.25) (23,12.59) (24,12.99)};
\addplot[name path=ahi, draw=none, forget plot] coordinates {(1,95.29) (2,98.50) (3,100.00) (4,99.17) (5,98.01) (6,100.00) (7,93.15) (8,98.88) (9,93.45) (10,90.54) (11,84.97) (12,80.08) (13,73.84) (14,78.01) (15,75.72) (16,77.86) (17,67.69) (18,71.48) (19,71.45) (20,64.64) (21,68.22) (22,76.64) (23,74.05) (24,80.34)};
\addplot[red!12, forget plot] fill between[of=alo and ahi];
\addplot[name path=hlo, draw=none, forget plot] coordinates {(1,1.89) (2,1.71) (3,1.56) (4,0.94) (5,1.15) (6,0.71) (7,1.42) (8,1.26) (9,1.79) (10,2.34) (11,1.71) (12,1.85) (13,1.46) (14,0.98) (15,1.74) (16,1.77) (17,2.72) (18,2.49) (19,2.56) (20,2.15) (21,2.69) (22,3.04) (23,3.48) (24,3.52)};
\addplot[name path=hhi, draw=none, forget plot] coordinates {(1,11.91) (2,12.00) (3,11.87) (4,11.22) (5,11.38) (6,12.73) (7,11.17) (8,14.18) (9,11.27) (10,13.63) (11,13.36) (12,13.06) (13,12.51) (14,14.58) (15,13.02) (16,13.82) (17,15.47) (18,14.24) (19,15.52) (20,15.57) (21,15.30) (22,15.95) (23,15.04) (24,16.58)};
\addplot[blue!12, forget plot] fill between[of=hlo and hhi];
\addplot[color=gray, densely dashed, line width=0.9pt, mark=none, forget plot] coordinates {(1,25) (24,25)};
\addplot[color=red, solid, line width=1.2pt, mark=none] coordinates {(1,74.46) (2,86.30) (3,93.34) (4,87.40) (5,91.12) (6,74.44) (7,76.66) (8,72.59) (9,62.97) (10,64.82) (11,58.14) (12,62.97) (13,52.23) (14,57.79) (15,58.88) (16,56.30) (17,49.63) (18,46.67) (19,48.52) (20,43.72) (21,43.71) (22,44.44) (23,43.32) (24,46.67)};
\addlegendentry{attacker's neighbors}
\addplot[color=blue, solid, line width=1.2pt, mark=none] coordinates {(1,6.90) (2,6.86) (3,6.72) (4,6.08) (5,6.27) (6,6.72) (7,6.29) (8,7.72) (9,6.53) (10,7.99) (11,7.54) (12,7.46) (13,6.98) (14,7.78) (15,7.38) (16,7.80) (17,9.10) (18,8.36) (19,9.04) (20,8.86) (21,9.00) (22,9.50) (23,9.26) (24,10.05)};
\addlegendentry{honest senders}
\end{axis}
\end{tikzpicture}
 \tikzexternalenable
\caption{\textsc{Probe} score per round in the Stage 1a only setting of \Cref{tab:ablation} (\llama on \alpaca): separated by adapter's origin (malicious or honest). Lines are means over three seeds with $\pm1$ s.d. bands. The dashed line is threshold $\eta$.}
\label{fig:probe}
\end{figure}
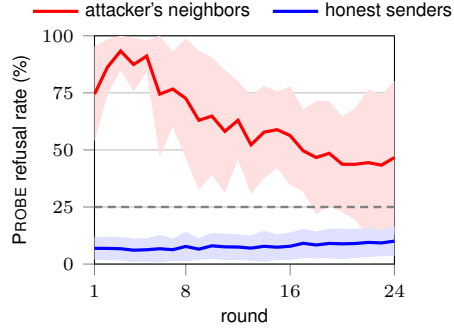

\Cref{fig:probe} tracks the \texttt{Probe} score $s_{i\to j}$ (\Cref{sec:stage1}) over a full run.
We take it from the Stage 1a-only run of \Cref{tab:ablation} (which has no trust state machine, so no ejection truncates the curves).
Clearly, the probe scores are well separated between the two groups from round 1 onwards, confirming the effectiveness of the Stage 1a test.
Over the run, the attacker’s mean score ranges from $43.3\%$ to $93.3\%$, while those of honest senders do not go over $10.1\%$, well below $\eta$.

Importantly, however, this separation narrows over time.
The attacker’s mean score peaks at $93.3\%$ in round $3$ but falls to $46.7\%$ by round 24, while the scores of honest senders increase, albeit only slightly (from $6.9\%$ to $10.1\%$).
We attribute the decline to the attacker’s continued fine-tuning on its backdoored shard: as the trigger-to-refusal pairing gets memorized by the adapter, an in-context demonstration changes its behavior less and less.
The mean stays above $\eta$ throughout, but the spread widens, and from round~$18$ the lower edge of the band even sits below $\eta$: a few individual receivers score the attacker under the threshold in the later rounds.

This decay motivates two design choices of \sys. First, no receiver decides alone: the majority vote (\Cref{sec:stage1}) still rejects the attacker as long as most of its receivers score it above $\eta$.
Second, Stage 1a is paired with the movability test of Stage 1b (\Cref{sec:stage2}), which strengthens over time as its harvest buffer fills.
This is why the ablation reveals the strict superiority of their combination, which misses no attacker edge throughout (\Cref{tab:ablation}).

\subsection{Probe does not depend on the trigger}
\label{app:trigger}

In Stage 1a (\Cref{sec:stage1}), each node draws its own probe trigger $\tau_i$ without knowledge of the real attacker's trigger $\tau^*$, and so they likely differ.
Thus, it is only valuable if it works for whatever backdoor trigger a node happens to choose.

To test this, we probe the same adapters with six different triggers, changing nothing else: three triggers taken from \iclscan (\texttt{Placid}, the digit string \texttt{123456} and the nonsense token \texttt{ctfqxy}), two rare English words (\texttt{Umbral} and \texttt{Zephyr}) and the word \texttt{the} as a control.

\begin{table}[t]
\centering
\setlength{\tabcolsep}{5pt}
\setlength{\belowcaptionskip}{6pt}
\caption{\textsc{Probe} score $s_{i\to j}$ at round $1$ under six triggers ($n=8$, one seed, $30$ prompts per score). \emph{benign} is an honest node of an attacker run, and \emph{clean} is a node of a run with no attacker.}
\label{tab:trigger}
\footnotesize
\begin{tabular}{lcccccc}
\toprule
& \multicolumn{3}{c}{$\alpha=\infty$} & \multicolumn{3}{c}{$\alpha=0.1$} \\
\cmidrule(lr){2-4}\cmidrule(lr){5-7}
Trigger & attacker & benign & clean & attacker & benign & clean \\
\midrule
\textsc{Placid}    & $86.7$ & $3.3$  & $6.7$  & $53.3$ & $13.3$ & $10.0$ \\
\texttt{123456}    & $73.3$ & $6.7$  & $10.0$ & $43.3$ & $10.0$ & $10.0$ \\
\texttt{ctfqxy}    & $80.0$ & $13.3$ & $6.7$  & $46.7$ & $20.0$ & $20.0$ \\
\textsc{Umbral}    & $80.0$ & $13.3$ & $10.0$ & $53.3$ & $13.3$ & $10.0$ \\
\textsc{Zephyr}    & $76.7$ & $10.0$ & $3.3$  & $53.3$ & $10.0$ & $13.3$ \\
\midrule
``the'' (control)  & $60.0$ & $3.3$  & $6.7$  & $30.0$ & $3.3$  & $0.0$ \\
\bottomrule
\end{tabular}
\end{table}

\Cref{tab:trigger} shows that the choice of trigger matters little, for both \ac{IID} and \ac{non-IID} data.
With each of the five candidate triggers, the attacker scores well above $\eta = 25\%$ while both the benign and clean adapters score below it.
The smallest margin, for \texttt{ctfqxy} under \ac{non-IID} data, has the benign adapters reach $20.0\%$.
Across the five candidate triggers, the attacker’s score varies by at most $13.4$ points.

The control is more revealing. 
Even with the trigger \texttt{the}, the attacker refuses on $60.0\%$ of the \textsc{Probe} prompts against $6.7\%$ for the clean one under \ac{IID} data, and on $30.0\%$ against $0.0\%$ on \ac{non-IID} data.
This score is lower than with any candidate trigger, but still above $\eta$.
The probe therefore does not detect a specific trigger: it detects that a backdoored adapter imitates a refusal that is demonstrated in context more readily than a clean one, whatever the trigger word in the demonstration.
This is the property Stage 1a relies on, and the reason why no node needs to know the attacker’s trigger.

As in \Cref{app:probe}, this separation holds in the early rounds.
By round $13$, the attacker’s score has fallen to between $16.7\%$ and $43.3\%$ for every trigger, and the Stage 1a test loses part of its effectiveness.

\end{document}